%% file: 0.sample-base.tex
\documentclass[sigconf]{acmart}

\usepackage{mdframed}
\usepackage{tabularx}
\usepackage{graphicx}
\usepackage{algorithm}
\usepackage{algpseudocode}

\AtBeginDocument{%
  }

\setcopyright{acmlicensed}
\copyrightyear{2025}
\acmYear{2025}
\acmDOI{XXXXXXX.XXXXXXX}

\acmConference[Conference acronym 'XX]{Make sure to enter the correct
  conference title from your rights confirmation emai}{June 03--05,
  2018}{Woodstock, NY}

\acmISBN{978-1-4503-XXXX-X/18/06}

\begin{document}



\title{In-Situ Mode: Generative AI-Driven Characters Transforming Art Engagement Through Anthropomorphic Narratives
}


\author{Yongming Li}
\authornote{The first authors.}
\affiliation{%
  \institution{Xi'an Jiaotong University}
  \city{Xi'an, Shaanxi}
  \country{China}}
\email{lym18733500195@stu.xjtu.edu.cn}

\author{Hangyue Zhang}
\affiliation{%
  \institution{University of Illinois Urbana-Champaign}
  \city{Champaign, Illinois}
  \country{USA}}
\email{hz85@illinois.edu}

\author{Andrea Yaoyun Cui}
\affiliation{%
  \institution{University of Illinois Urbana-Champaign}
  \city{Champaign, Illinois}
  \country{USA}}
\email{yaoyunc2@illinois.edu}

\author{Zisong Ma}
\affiliation{%
  \institution{University of Illinois Urbana-Champaign}
  \city{Champaign, Illinois}
  \country{USA}}
\email{zisongm2@illinois.edu}

\author{Yunpeng Song}
\affiliation{%
  \institution{Xi'an Jiaotong University}
  \city{Xi'an, Shaanxi}
  \country{China}}
\email{sypxjtu@gmail.com}

\author{Zhongmin Cai}
\authornote{Both authors are corresponding authors.}
\affiliation{%
  \institution{Xi'an Jiaotong University}
  \city{Xi'an, Shaanxi}
  \country{China}}
\email{zmcai@sei.xjtu.edu.cn}

\author{Yun Huang}
\authornotemark[2]
\affiliation{%
  \institution{University of Illinois Urbana-Champaign}
  \city{Champaign, Illinois}
  \country{USA}}
\email{yunhuang@illinois.edu}

\renewcommand{\shortauthors}{Li et al.}

\begin{abstract}

Art appreciation serves as a crucial medium for emotional communication and sociocultural dialogue. In the digital era, fostering deep user engagement on online art appreciation platforms remains a challenge. Leveraging generative AI technologies, we present \textit{EyeSee}, a system designed to engage users through anthropomorphic characters. We implemented and evaluated three modes-- \textcolor{blue}{\textit{Narrator}}, \textcolor{orange}{\textit{Artist}}, and \textcolor{green}{\textit{In-Situ}}--acting as a third-person narrator, a first-person creator, and first-person created objects, respectively, across two sessions: \textit{Narrative} and \textit{Recommendation}. We conducted a within-subject study with 24 participants. In the \textit{Narrative session}, we found that the \textit{In-Situ} and \textit{Artist modes} had higher aesthetic appeal than the \textit{Narrator mode}, although the \textit{Artist mode }showed lower perceived usability. Additionally, from the \textit{Narrative} to \textit{Recommendation session}, we found that user-perceived relatability and believability within each interaction mode were sustained, but the user-perceived consistency and stereotypicality changed. Our findings suggest novel implications for applying anthropomorphic in-situ narratives to other educational settings.

\end{abstract}



\begin{teaserfigure}
    \centering
    \includegraphics[width=1\linewidth]{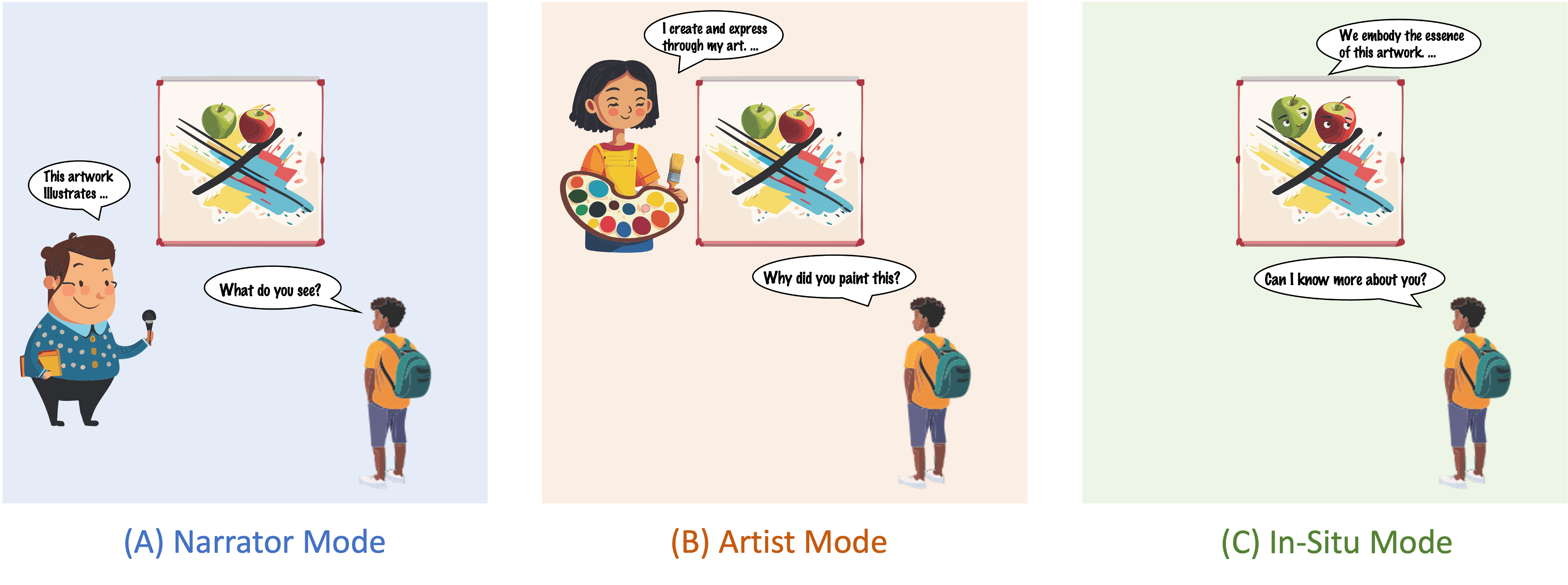}
    \caption{\texttt{The \textit{EyeSee} system introduces three interaction modes, designed to enhance art appreciation through anthropomorphic narratives. The  \textcolor{blue}{\textit{Narrator mode}} (panel (A)) functions as a third-person storyteller, providing users with contextual information and background about the artwork. The \textcolor{orange}{\textit{Artist mode}} (panel (B)) adopts the perspective of the artwork's creator, offering insights into the artistic process and motivations. The \textcolor{green}{\textit{In-Situ mode}} (panel (C)) presents the viewpoint of an object or figure within the artwork, allowing users to engage with the narrative from an internal perspective. Note that no modifications were made to the original artwork; the eyes on the apple in the third panel are included for illustrative purposes only.}}
    \Description{The left picture shows a tour guide explaining the artwork to a visitor. In the middle picture, a woman resembling an artist is providing an explanation of the painting to the same visitor. In the picture on the right, there are no other people present except for the visitor and the artwork itself. The objects within the painting come to life and begin explaining the piece to the visitor on their own.}
    \label{fig:teaser}
\end{teaserfigure}



\maketitle

\input{SECTIONS/1.Introduction}

\input{SECTIONS/2.LiteratureReview}

\input{SECTIONS/4.2.EyeSeeSystem}

\input{SECTIONS/5.UserStudy}


\input{SECTIONS/6.1.Findings1-1}

\input{SECTIONS/6.1.Findings1-2}

\input{SECTIONS/6.1.Findings1-3}

\input{SECTIONS/6.1.Findings1-4}

\input{SECTIONS/6.2.Findings2-1}

\input{SECTIONS/6.2.Findings2-2}

\input{SECTIONS/7.Discussion}

\begin{acks}
This project was made possible in part by the Institute of Museum and Library Services RE-252329-OLS-22. The views, findings, conclusions or recommendations expressed in this article do not necessarily represent those of the Institute of Museum and Library Services.
\end{acks}

\bibliographystyle{ACM-Reference-Format}
\bibliography{0.sample-base}


\end{document}

%% file: SECTIONS/1.Introduction.tex
\section{INTRODUCTION}

\begin{quote}
    \centering
    \textit{"Art asks us to think differently, see differently, hear differently, and ultimately to act differently, which is why art has moral force." --Jeanette Winterson~\cite{sherman2017art} }
\end{quote}

Art has long been recognized as a critical medium for expressing emotions, communicating ideas, and reflecting on cultural and personal experiences~\cite{dervin2023critical,silva2006distinction}. \textit{Art appreciation} entails the viewer's active engagement, as the meaning and value of visual art are constructed through the interactions and conversations between the artwork, the artist, and the viewer~\cite{holub2013reception, kemp1998work, lee2024llava}. The appreciation of visual art makes us think about ‘who we are’, ‘how we interact with others’, and ‘our place in society’, thus it serves a crucial role not only in enriching personal life but also in reflecting societal values and fostering a deeper understanding of cultural differences~\cite{smith2014museum}. Analyzing complex visual arts contributes to intellectual and emotional growth and discussions about art often extend to broader conversations about ethics and values~\cite{barnard1998art,acharya2023applicability}. 
However, engaging people with art in an era dominated by digital experiences remains a challenge. Though traditional online art appreciation platforms have made art more accessible to a broader audience by overcoming geographical limitations and reducing costs through virtual tours~\cite{ilic2019information, lee2024llava}, they often lack interactivity~\cite{sa2024strategy, shi2024reconstructing}, resulting in a passive viewing experience that fails to offer the depth of engagement necessary for a profound appreciation of visual art. This gap highlights the importance of enhancing the interactivity and engagement levels of online art platforms~\cite{morse2022museumathome, macdonald2015assessing}.

In response to this challenge, the development of generative AI technologies offers new opportunities to enhance user engagement by simulating anthropomorphic characters~\cite{qin2024charactermeet, salminen2024deus, liu2024personality} and providing more personalized interactive experiences ~\cite{tao2024does,  hayashi2024artwork}. For example, researchers found that LLMs improved the ability of non-experts to discern truth in debates by simulating diverse expert characters~\cite{khan2024debating}, and diverse LLM-simulated student characters helped students practice mathematical modeling skills in educational settings~\cite{yue2024mathvc}. 
In the context of art appreciation, AI-powered chatbots can simulate a third-person docent character to generate reflective questions about paintings, thereby helping users engage more deeply with the artwork~\cite{gollapalli2023generating}. Similarly, Lee et al.~\cite{lee2024llava} simulated a personal tutor for art appreciation to provide personalized support and enhance users' comprehension of the artworks. However, despite these advancements, there is still a need to explore how different character perspectives, such as those of created objects in a painting and the painting's creator, can promote sustained user engagement, particularly in open art appreciation contexts.

In this paper, we present \textit{EyeSee}, designed to address this gap by introducing multi-character interactions that engage users through anthropomorphism. We designed, implemented, and evaluated three distinct modes: the \textcolor{blue}{\textit{Narrator}}, the \textcolor{orange}{\textit{Artist}}, and the \textcolor{green}{\textit{In-Situ}} modes. The \textit{Narrator} provides objective explanations of the artwork, the \textit{Artist} offers insights from the creator's perspective, and the \textit{In-Situ} mode enables interaction with the created objects or figures within the artwork. This approach is inspired by “characters built with contextual data”\cite{miaskiewicz2011personas}, which involves creating characters based on contextual information, and the concept of "thing-centered narratives"\cite{cila2015thing}, which focuses on narratives centered around objects.

To explore how different interaction modes affect users' aesthetic appeal and immersive experiences, we conducted a within-subject study with 24 participants, where each participant experienced all three modes, each offering both \textit{Narrative} and \textit{Recommendation} features during interactions. In the \textit{Narrative} \textit{session},
the participants engaged with the narratives about painting objects and we found that both the \textit{Artist} and \textit{In-Situ} modes elicited higher aesthetic appeal, which was reflected in increased emotional engagement, particularly related to themes such as time travel, empathy, and anthropomorphism. The use of first-person perspectives in these modes might have contributed to this effect. However, the usability of the \textit{Artist} mode was perceived as lower, possibly due to higher knowledge expectations and stricter accuracy demands that diminished cognitive engagement. Transitioning from the \textit{Narrative} to the \textit{Recommendation} \textit{session}, the \textit{In-Situ} mode consistently scored highest in relatability and believability. Moreover, compared to the \textit{Narrative} \textit{session}, the \textit{Artist} mode exhibited improved perceived response consistency during the \textit{Recommendation} \textit{session}. This can be attributed to the artist character's suitability for making contextual recommendations, thus enhancing perceived epistemic value.

Our work makes novel and significant contributions to the HCI field. \textit{First}, our study demonstrates the superior performance of \textit{In-Situ} design over traditional narrative formats in enhancing user engagement. The \textit{In-Situ} design excels across various metrics including focused attention, usability, aesthetic appeal, and reward. These improvements underscore the role of anthropomorphism and contextual narratives in fostering deeper connections with art. 
\textit{Second}, our empirical findings reveal that employing multi-perspective interactions, especially through first-person narratives, significantly boosts emotional engagement with visual art, enriching the user's art appreciation experience.
\textit{Third}, our study demonstrates different interaction modes and user engagement levels, particularly cognitive and emotional engagement in interactive \textit{Narrative session}, significantly influence satisfaction of the recommended content and perception of recommendation reasons provided by character (i.e. relatability, and believability).
\textit{Fourth}, we explore the application of our approach, driven by a Multimodal Large Language Model (MLLM), in educational contexts. Here, multi-perspective strategies could foster more meaningful engagement with arts and other humanities disciplines, thus enhancing educational experiences. 


%% file: SECTIONS/2.LiteratureReview.tex
\section{Related Work}

\subsection{Art Appreciation and Interactive Engagement}

Art appreciation enriches personal life and fosters cultural reflection and sociocultural communication. Engaging with art, even in short art appreciation sessions, has been shown to offer mental and physical benefits, such as reducing stress, improving mood~\cite{ho2015art}, and lowering cortisol levels~\cite{clow2006normalisation} and blood pressure~\cite{mastandrea2019visits}. Beyond these immediate health benefits, art appreciation allows individuals to immerse themselves in diverse historical contexts and viewpoints, fostering personal reflection and self-expression~\cite{belfiore2007determinants, sherman2017art, durlak2011impact}. The emotional and cognitive engagement required in art appreciation evokes profound emotional responses and stimulates introspective thought~\cite{sherman2017art}. Furthermore, at the societal level, art serves as a medium for discourse~\cite{dervin2023critical,silva2006distinction}.  Communities form around shared artistic interests, leading to discussions that extend to broader topics such as ethics, values, and societal norms~\cite{gaztambide2008artist, van2009study}. These discussions help foster a deeper understanding of cultural differences and promote inclusivity~\cite{cole1998can, sherman2017art}.

Digital platforms have transformed the art appreciation process, enabling more profound and contemplative engagement with artworks~\cite{walmsley2016arts}. These platforms have addressed traditional barriers such as geographical limitations and cost by offering virtual tours and high-resolution images of art collections, thus making art more accessible to a broader audience~\cite{ilic2019information}. Beyond accessibility, digital platforms have contributed significantly to art education. For example, artificial intelligence, particularly deep learning, has been used to assist students in understanding and categorizing artworks, thereby enriching their educational experiences and enjoyment~\cite{chiu2024artificial, yi2022driis, hung2022learning}.  In the domain of online platforms, interactive engagement plays a crucial role in enhancing the appreciation experience~\cite{shen2009affective}. By integrating elements such as virtual reality and conversation agents, online platforms transform passive viewers into active participants~\cite{moon2024mixed, scholz2016augmented, yuen2011augmented, wang2024virtuwander}.

\subsection{LLM-enabled Anthropomorphism for Different Role-plays}

Anthropomorphism refers to the psychological phenomenon of “attributing human characteristics to the nonhuman entities”~\cite{seeger2018designing}. In AI systems, anthropomorphic design can significantly influence user expectations, trust, and interaction quality\cite{jensen2021disentangling, ma2023understanding, wu2024sunnie}. 
Design features of anthropomorphic characters are generally categorized into social cues and verbal cues. Social cues, such as the use of text-to-speech voices, have been shown to enhance the perceived anthropomorphism of conversational interfaces compared to text-only interactions~\cite{cohn2024believing, moussawi2021perceptions}. Verbal cues, like the use of first-person pronouns ("I"), have been found to increase perceived information accuracy and reduce risk in specific contexts, such as medication counseling~\cite{cohn2024believing}.

Large Language Models (LLMs) have demonstrated remarkable capabilities in generating anthropomorphism characters to improve user experience. 
For example, in the mental health context, Louie et al.~\cite{louie2024roleplay} simulated patient characters to help novice counselors practice their social skills. They found that a novel principle-adherence prompting pipeline improved response quality and adherence to expert-defined principles by 30\%. In education, LLMs have been employed to embody various anthropomorphic characters to enhance the accuracy and professionalism of their generated text~\cite{hu_quantifying_2024,louie_roleplay-doh_2024}. For example, when assigned specific characters like historians or scientists, LLMs produced more precise and domain-specific responses, enhancing both creativity and accuracy~\cite{hu_quantifying_2024, lu_llm_2024}. Similarly, Arguedas and Daradoumis~\cite{arguedas2021analysing} demonstrated that a pedagogical tutor character providing cognitive and affective feedback positively influenced students' perceptions by stimulating engagement and guiding learning activities. In the realm of art appreciation, Lee et al.~\cite{lee2024llava} applied LLMs to simulate student and teacher characters, leading to the development of the LLaVA-Docent. This multimodal large language model was designed as a personal tutor for art appreciation, providing interactive, engaging experiences that support deeper learning and engagement with artworks. These studies signify a shift from the traditional, one-size-fits-all generic agent character to more personalized and specialized AI-enabled characters, tailored to enhance user engagement across different domains.

\subsection{Multiple Characters' Perspectives in Art Appreciation Context}

The narrative perspective plays a crucial role in how users connect with the anthropomorphic character. Different perspectives can substantially enhance engagement and user experience in narrative contexts~\cite{chen2024persona,chen2021changing}. While there is ongoing debate among scholars about the fundamental differences between first-person and third-person perspectives~\cite{stanzel1984theory,kaufman2012changing}, both perspectives can enhance user experience and emotional connection in specific contexts. Specifically, third-person perspectives were found helpful in increasing user trust in characters~\cite{van2016difficult} and helping readers understand characters' actions and thoughts~\cite{al2019point}. However, because the characters' thoughts and feelings are described from an anonymous external viewpoint, they seem more distant and abstract to readers~\cite{kim2020presence}. Compared to third-person perspectives, adopting a first-person perspective often elicits greater narrative engagement~\cite{busselle2009measuring}. For example, Salem et al.~\cite{salem2017does} showed that first-person narration boosts the connection to the protagonist. Another study by Samur et al.~\cite{samur2021getting} proved that first-person stories elicit greater narrative engagement compared to third-person stories. Brennan~\cite{brennan2024versus} proposed that writing research articles in the first person made them more engaging, creative, and interesting for readers. Additionally, personal experiences, cultural backgrounds, and emotional states shaped users’ subjective interpretations of art~\cite{belfiore2007determinants}, all of which emphasized the need to incorporate diverse perspectives~\cite{hooper2000changing} in online platforms integrating. Generative AI present a promising approach to broadening narrative perspectives, thereby enriching users' engagement and appreciation of art.

%% file: SECTIONS/4.2.EyeSeeSystem.tex
\section{\textit{EyeSee} DESIGN AND IMPLEMENTATION}

In this section, we introduced \textit{EyeSee}, a multi-character prototype designed to explore \textbf{how users perceive and engage with anthropomorphic characters across three modes—\textcolor{blue}{\textit{Narrator}}, \textcolor{orange}{\textit{Artist}}, and \textcolor{green}{\textit{In-Situ}}—differently, and why}. Unlike previous studies that focused on single-perspective chatbots, \textit{EyeSee} incorporates both first-person and third-person perspectives, providing a more comprehensive analysis of user engagement, pleasure, and knowledge gained. Previous research showed that the first-person perspective is associated with higher emotional and cognitive engagement~\cite{van2016difficult}. In particular, first-person narratives have been found to generate emotional engagement and prompt behavioral intent compared to the third-person perspective~\cite{liu2024analysis}. To leverage these benefits, we applied this perspective to two modes: the \textit{Artist} and the \textit{In-Situ} modes.

The \textbf{\textit{Artist mode}} was designed to interpret the artwork from the creator's perspective. Drawing on Bullot et al.~\cite{bullot2013artful}, who emphasized that an understanding of the artist’s background and creative motives could enhance viewer appreciation, this mode sought to provide information from the \textit{Artist} \textit{character} perspective. 
The \textbf{\textit{In-Situ mode}} presented the viewpoint of an object or figure within the artwork, allowing users to engage with the narrative from an internal perspective. This mode was built on Cila et al.~\cite{cila2015thing}, who introduced "thing-centered narratives," a concept demonstrating how objects could convey human-like information and offer new viewpoints on familiar practices. Recent work in the HCI field, such as Coskun et al.~\cite{coskun2022more}, also confirmed the potential of super-human design perspectives.
Finally, the \textbf{\textit{Narrator mode }}employed a third-person perspective, similar to traditional museum guides~\cite{hein2002learning,best2012making}. This mode provided objective explanations and contextual information about the artwork.

To reflect common use cases in online art appreciation platforms and to investigate \textbf{how user perception evolves across different task sessions, and why}, we designed two task sessions: the \textit{Narrative} \textit{session} and the \textit{Recommendation} \textit{session}.  Current platforms like Google Arts \& Culture and virtual museums primarily relied on static images and textual descriptions, which limited user engagement and interaction~\cite{sa2024strategy, shi2024reconstructing}. Also, users often faced challenges when using tools like GPT to locate specific objects in artworks. The \textbf{\textit{Narrative session}} allowed users to explore specific areas of interest within artworks. Current art recommendation systems often failed to fully utilize user interaction data and lacked transparency. \textbf{\textit{Recommendation session}} addressed this by providing personalized recommendations based on minimal user interaction, accompanied by explanations from multiple perspectives.

\subsection{\textbf{\textit{EyeSee} Interaction Design}}
\

In this session, we present the final version of the \textit{EyeSee} system. The \textit{EyeSee} system includes one study setup interface and two task interfaces: \textit{Narrative} interface and \textit{Recommendation} interface. 

\begin{figure*}[t!]
\centering
\includegraphics[width=0.85\textwidth]{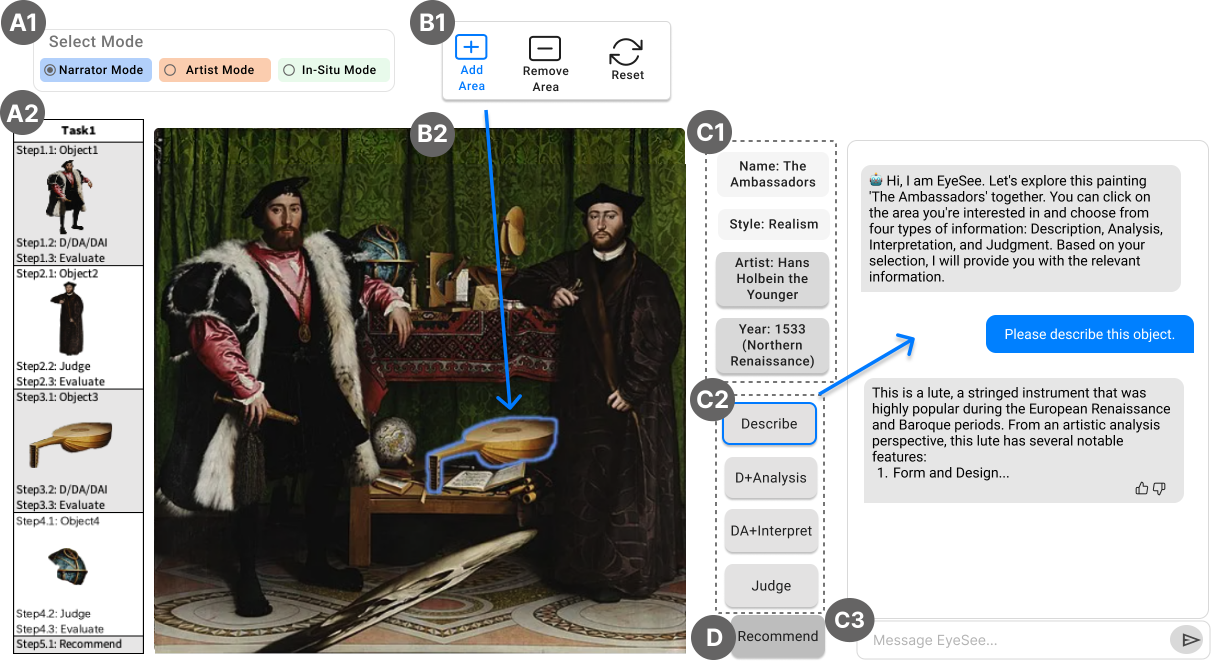}
\caption{\texttt{\textit{Narrative} Interface Include (A1) AI Agent Character Selection Area, Including Narrator, Artist, and In-Situ Modes; (A2) Task Instruction Panel; (B1) Area of Interest Selection, with Buttons to Add, Remove, or Reset Selection Areas; (B2) Attention Area Display; (C1) Basic Metadata Information: Name, Style, Artist, and Year; (C2) Shortcuts for Art Appreciation Information: Describe, Describe + Analysis, Describe + Analysis + Interpret, and Judge; (C3) Free Question Query; and (D) Artwork Recommendations.}}
\Description{This screenshot showcases the user interface of the EyeSee narration system while exploring the artwork "The Ambassadors." The artwork is positioned in the center of the screen. On the left side, a series of five interactive tasks are listed vertically, each corresponding to different objects within the artwork for user engagement. Above the artwork, a toolbar allows users to select different modes of interaction. Next to this toolbar, there is an additional panel where users can zoom in or out on selected areas or objects within the artwork, or clear their selection. In this case, the lute in the artwork is highlighted in blue, indicating that the user has chosen this object for the current interaction. To the right of the artwork, detailed information about the piece is displayed, including the title, artist, year, and style. Below this information, several vertically arranged buttons allow users to perform actions such as describing, analyzing, or interpreting specific elements of the painting. In this screenshot, the “Describe” mode is active, as indicated by the selection in the system, and the system’s response regarding the lute is shown in the chatbox on the far right, providing information about the object’s historical significance and design.}
\label{narration}
\end{figure*}

First, as shown in Figure \ref{narration}, the components (A1) and (A2) are designed for experiment setup: (A1) allows users to set the AI agent mode, offering a choice between the \textit{Narrator}, the \textit{Artist}, and the \textit{In-Situ modes}, and (A2) provides step-by-step task instructions to help users understand the required actions in two task sessions. 

\begin{figure*}[t!]
\centering
\includegraphics[width=0.85\textwidth]{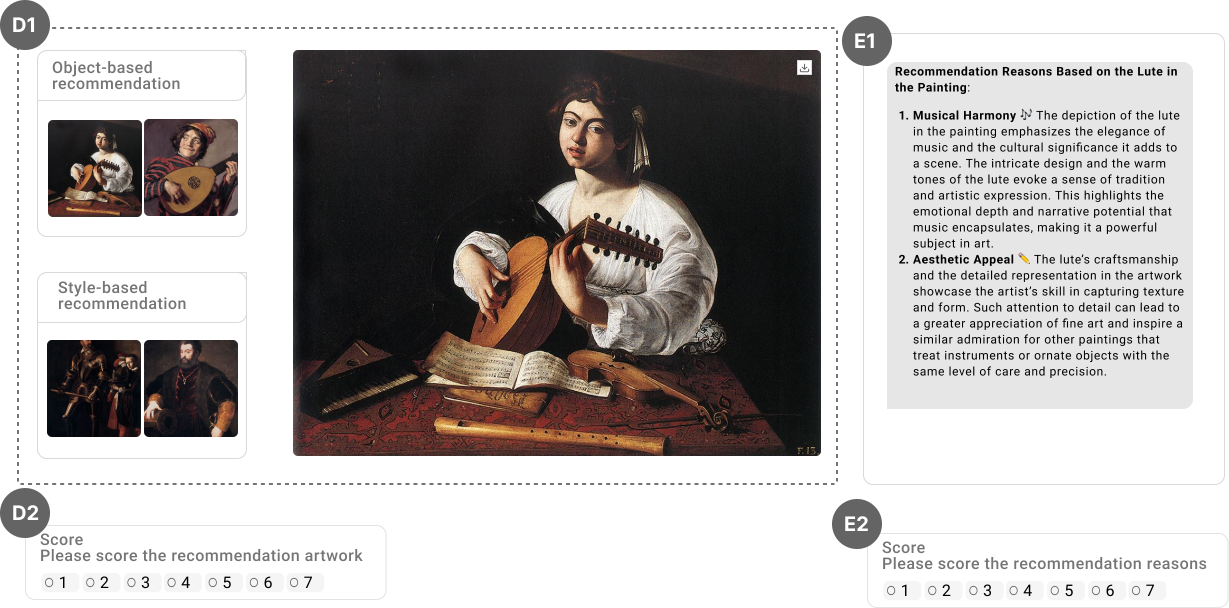}
\caption{\texttt{\textit{Recommendation} Interface Include (D1) Recommended Artwork Display; (D2) Rating for Recommended Artworks; (E1) Recommendation Reasons; and (E2) Rating for Recommendation Reasons.}}
\label{recomandation}
\Description{This screenshot showcases the user interface of the EyeSee recommendation system after the user selected the lute in the previous artwork, The Ambassadors. As a result, the system generated recommendations, and the user is currently viewing the first recommended artwork, The Lute Player by Caravaggio, which is displayed prominently in the center of the screen. On the left side (D1), there are four thumbnail images representing the recommended artworks. The top two thumbnails are based on object similarity, specifically featuring a lute, while the bottom two are holistic recommendations, suggesting artworks that share a similar overall style to The Ambassadors. The user can click on any of these thumbnails to view them in detail. On the right side (E1), the system presents the reasons for recommending The Lute Player, while below (E2), the user is invited to rate the reasons for the recommendation on a scale of 1 to 7. Similarly, at the bottom left (D2), the user is prompted to rate the recommended artwork itself. }
\end{figure*}
\

Second, the components (B1) and (B2) are designed for the attention area selection. In the B1 area, the three buttons function as follows: "Add Area" allows users to add clicked areas to the selected attention area, "Remove Area" subtracts clicked areas from the selected attention area, and "Reset" clears all attention areas. In the B2, users can upload paintings and see the selected attention areas displayed in real-time. The components (C1), (C2), and (C3) are designed for information type selection. C1 provides the name, style, artist, and year of the artwork. C2 draws on 
Feldman's Model of Art Criticism~\cite{feldman1994practical} to offer shortcuts about the description, analysis, interpretation, and judgment of selected attention areas. C3 allows users to ask questions freely. The system offers the above information by interactive dialogue based on the selected character. The component D guides users to the EyeSee recommendation interface, where personalized artwork recommendations are provided based on the user’s selected attention areas and preferences.

Third, as shown in Figure \ref{recomandation}, the components (D1) and (D2) are designed for displaying and evaluating recommended artworks. D1 provides four personalized recommendations: 
the first two paintings are recommended based on the user's interest in the painting style, while the last two are based on the user's selected areas of interest during the \textit{Narrative} \textit{session}. Users can click on paintings to preview them. D2 is used to collect satisfaction ratings for the recommended paintings. In the deployed system, D2 enables users to bookmark paintings that interest them, making it easy to revisit the collection. The components (E1) and (E2) are designed for displaying and evaluating the reasons behind the recommendations. E1 displays the recommendation reason when users click on an image in the preview area. The reason is based on the relationship between the recommended artwork and the original artwork using the same character perspective (\textit{Narrator, Artist, or In-Situ}) as in the \textit{Narrative} \textit{session}. E2 allows users to rate the recommendation reasons on a scale from 1 to 7, where 1 represents “very dissatisfied” and 7 represents “very satisfied”.

\subsection{\textit{EyeSee} Backend and Implementation}

\begin{figure*}
\centering
\includegraphics[width=0.85\textwidth]{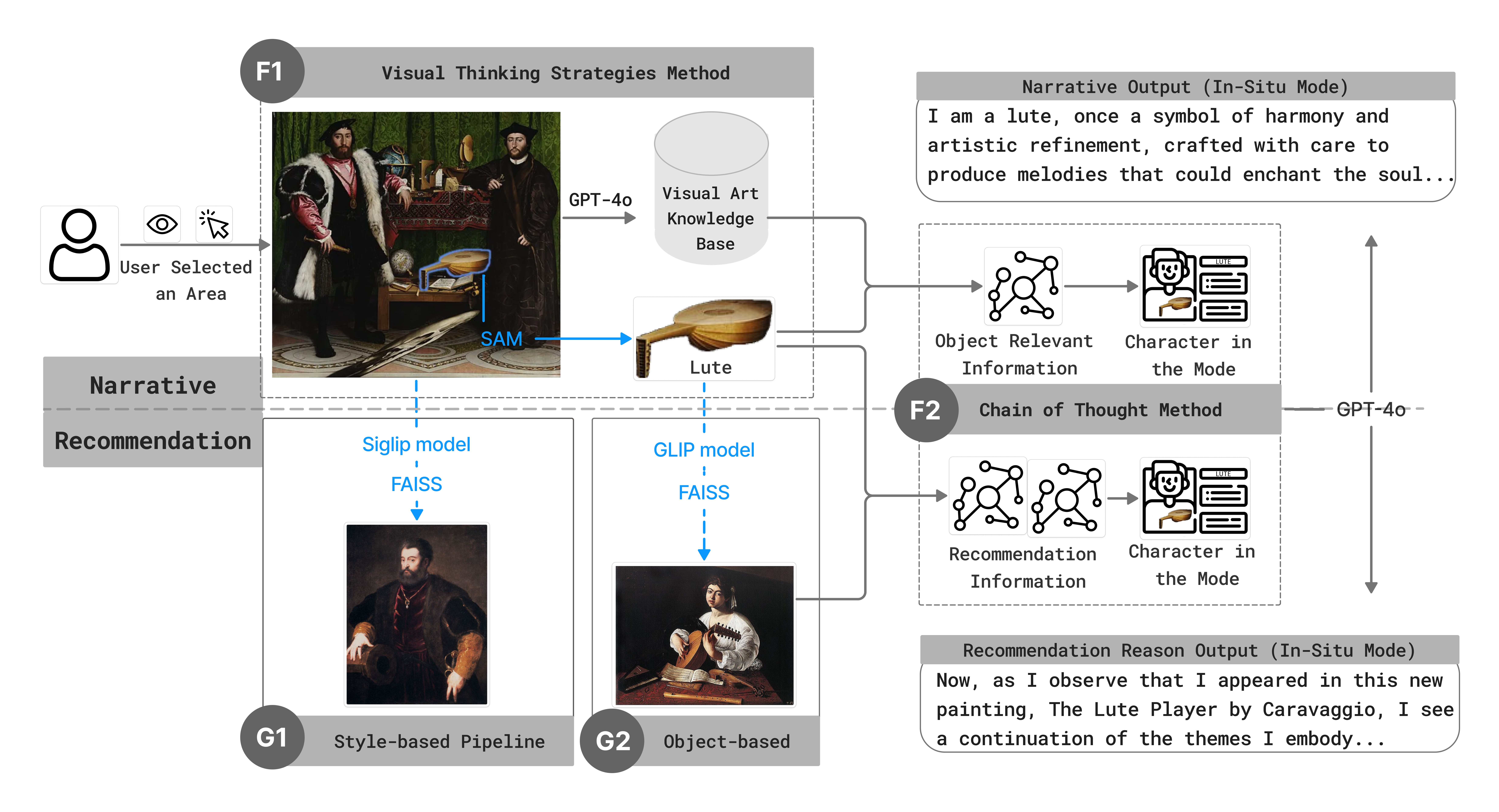}
\caption{\texttt{EyeSee framework include: (F1) Visual Thinking Strategies Method; (F2) Chain of Thought Method; (G1) Style-based Pipeline; (G2)Object-based Pipeline.}}
\Description{This flowchart illustrates the backend process of the EyeSee system. The process begins when the user selects an area of interest in the artwork. The system, using the VTS Method (F1), identifies the selected object, such as the lute, and queries the visual art knowledge base through GPT-4. The system generates a narrative output based on the selected object, reflecting its historical and artistic context. Simultaneously, the system applies both a Style-based Pipeline (G1) and an Object-based Pipeline (G2), using machine learning models (like FAISS) to recommend artworks similar in style or containing similar objects. The recommendations are then processed through the CoT Method (F2), which combines object-relevant and recommendation information to generate personalized outputs that provide both narrative and recommendation insights for the user.}
\label{backend}
\end{figure*}

The backend of the \textit{EyeSee} system utilizes three modes-- \textit{Narrator}, \textit{Artist}, and \textit{In-Situ}--to customize MLLM-based agents for generating narratives and recommendations based on users' areas of interest. Two major functions of the \textit{EyeSee} backend include: 1) the narrative generation ability of the MLLM agent and 2) the artworks retrieval module that supports artwork recommendation. The framework of \textit{EyeSee} system is shown in Figure \ref{backend}.

\textbf{Generating Narrative with MLLM-based Agent.}
The \textit{EyeSee} system leverages the Visual Thinking Strategies (VTS) method (F1) to build the visual art knowledge base and the Chain of Thought (CoT) method (F2) to customize the MLLM-based agent to different characters for generating narratives. Developed at the Museum of Modern Art in New York City, Visual Thinking Strategies~\cite{yenawine2013visual} have been widely adopted in art appreciation education for facilitating participants to express their interpretations of the artwork. This method guides the GPT-4o model in generating a visual art knowledge base. This method relies on three open-ended questions, that is "What’s going on in this picture?", "What do you see that makes you say that?", and "What more can you find?", to build the visual art knowledge base. 
Secondly, drawing upon the proven efficacy of chain-of-thought’s applications in diverse fields \cite{kim-etal-2023-cotever, si2023getting, feng2024towards}, the \textit{EyeSee} system incorporates the Chain of Thought method to generate character-based narratives through intermediate steps. This process involves three steps: (1) extracting and identifying the objects based on the Segment Anything Model~\cite{kirillov2023segment} (a groundbreaking image segmentation algorithm) and labels (some named entities extracted from the knowledge base); (2) extracting relevant statements from the knowledge base based on the object's name and information types; (3) transforming these relevant statements into narratives based on the perspective of the chosen character. Examples of prompts used to generate narratives are available in the Appendix.

\textbf{Related Artworks Retrieval.}
The \textit{EyeSee} system employs two retrieval pipelines to recommend existing artworks based on the user’s interest in painting styles and selected objects.
First, the style-based retrieval pipeline (G1) is designed to recommend artworks that match the users' preferred painting style by comparing the similarity of the image feature vector. The \textit{EyeSee} system utilizes the Siglip model (siglip-base-patch16-224)~\cite{zhai2023sigmoid} to extract features from the painting, with these vectors capturing the essence of the artwork’s style.  Then, the system uses FAISS~\cite{johnson2019billion}, a fast similarity search tool, to retrieve two paintings with the most similar features from the  Wikiart datasets~\cite{danielczuk2019segmenting}, ensuring the recommendations align with the user’s stylistic preferences.
Second, the object-based retrieval pipeline (G2) recommends artworks containing user-selected objects using image segmentation and recognition models. The system follows the object detection pipeline proposed by Louie Meyer et al. \cite{10.1145/3613904.3642157}. with the Wikiart datasets are annotated using 13 categories and 120 labels. Specifically,  during the data annotation process, the \textit{EyeSee} system utilizes the pre-trained GLIP model (glip-tiny-model-o365-goldg)~\cite{li2022grounded} to compare the vector representation of an object label with the vectors extracted from image patches, matching the most similar ones and generating labeled bounding boxes within the painting. Then, the system uses the Segment Anything Model~\cite{kirillov2023segment} to extract objects from paintings. It then applies FAISS~\cite{johnson2019billion} to compare the vector representation of the selected object with those in the Wikiart dataset~\cite{danielczuk2019segmenting}, retrieving two similar paintings based on similarity scores.  Additionally, the \textit{EyeSee} system also incorporates the Chain of Thought method into recommendation reason generation: (1) extracting and identifying the object; (2) analyzing the recommendation reasons based on two paintings' knowledge base; (3) transforming these relevant statements into recommendation reasons.

\textbf{Iterative Design and Implementation.} The design and implementation of the \textit{EyeSee} system were carried out by three authors of this paper between March to June 2024, following the Action Design Research~\cite{hansen2019participatory,mullarkey2019elaborated}, which includes diagnosis, design, implementation, and evolution phases. The system underwent five iterative design cycles, each guided by usability testing and technical considerations.

%% file: SECTIONS/5.UserStudy.tex
\section{Method}

\input{SECTIONS/8.Appendix8-3}

\begin{figure*}
\centering
\includegraphics[width=0.9\textwidth]{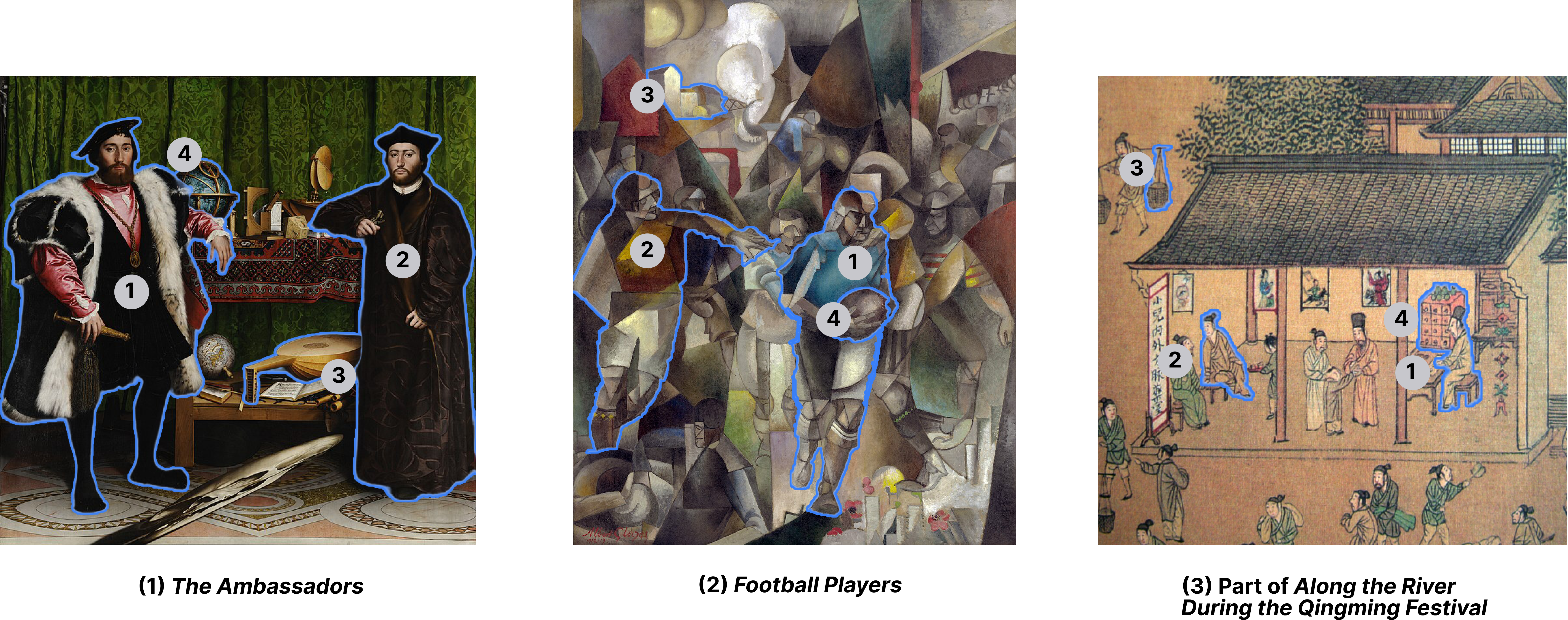}
\caption{\texttt{Experiment Materials}}
\label{Materials}
\Description{The diagram illustrates experiment materials including three paintings, organized from left to right: (1) \textit{The Ambassadors} (Hans Holbein the Younger, 1533, Northern Renaissance); (2) \textit{Football Player} (Albert Gleizes, 1912-1913, Cubism); and (3) \textit{Along the River During the Qingming Festival} (Zhang Zeduan, Song Dynasty, Chinese Landscape Painting).}
\end{figure*}

\begin{figure*}
\centering
\includegraphics[width=1\textwidth]{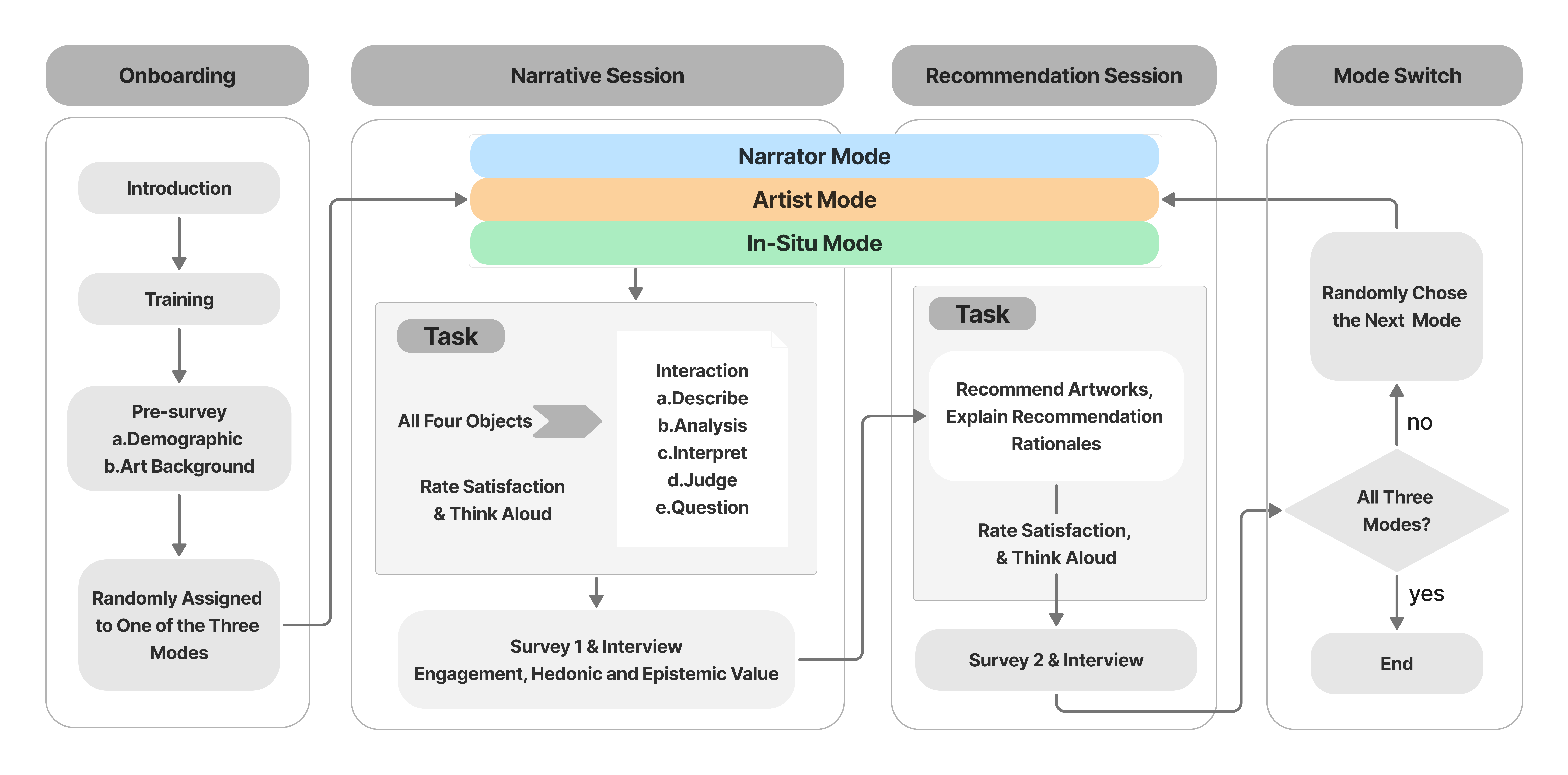}
\caption{\texttt{Experiment Procedure Includes Onboarding Session, \textit{Narrative Session}, and \textit{Recommendation Session}}}
\label{Study Process}
\Description{This flowchart outlines the process of using the EyeSee system, starting with the Onboarding phase, where users go through an introduction, training, and a pre-survey, and are then randomly assigned to one of three modes: Narrator, Artist, or In-Situ. In the Narrative Session, users interact with four objects, performing tasks like describing, analyzing, interpreting, and answering questions. Following this, they rate their satisfaction and complete a survey. In the Recommendation Session, users are recommended artworks and given rationales, after which they again rate satisfaction and complete a survey. Finally, the Mode Switch phase allows users to switch modes randomly until all three modes have been explored, after which the process ends. }
\end{figure*}

\subsection{Participants}
\

As shown in Table \ref{tab:participanttable}, 24 participants participated in the experiment (age Mean = 23.63, SD = 3.84; 14 identified as female, 8 as male, and 2 as other). Participants were recruited through electronic flyers and emails, using snowball sampling to target individuals with a demonstrated interest in art appreciation, as art enthusiasts were more likely to provide meaningful and insightful feedback. The participants represented a diverse range of backgrounds, including 5 graduate students, 11 undergraduates, 2 recent high school graduates, 2 art teachers, 1 freelance artist, 1 graphic designer, 1 art archaeology researcher, and 1 museum staff member. The experiment received approval from the Institutional Review Board (IRB) at the researchers' institution. All participants signed the informed consent form. and were compensated at a rate of US\$20 per hour upon completing the experiment.

\subsection{Experiment Materials and Tools}

As shown in Figure \ref{Materials}, we used the three paintings in the experiment: (1) \textit{The Ambassadors} (Hans Holbein the Younger, 1533, Realism); (2) \textit{Football Player} (Albert Gleizes, 1912-1913, Cubism); and (3) \textit{Along the River During the Qingming Festival }(Zhang Zeduan, Song Dynasty, Chinese Landscape Painting). The paintings were chosen to ensure the experiment materials include various styles and periods. Additionally, we selected four objects in each painting: two human figures and two non-human objects for the object-based interaction. These paintings were obtained from three online museum websites\footnote{https://www.nationalgallery.org.uk/} \footnote{https://www.nga.gov/} \footnote{https://www.comuseum.com/}.

The experiment was conducted remotely using the following tools: (1) a laptop, a mouse, and earphones when the participant visited the online \textit{EyeSee} system. (2) a PC and remote meeting recording software used to collect participant’s interaction logs and feedback. In addition, the scales used in the experiment were created in Qualtrics, and the collected data were analyzed in Python.

\subsection{Experiment Procedure}

As shown in Figure \ref{Study Process}, the experiment procedure consists three sessions: (1) an onboarding session where participants received instructions, familiarized themselves with the prototype, and completed pre-survey; (2) a \textit{Narrative session} where participants interacted with the three modes of \textit{EyeSee} to obtain information about the selected objects and evaluate engagement during the narrative task; (3) a \textit{Recommendation session} where participants rated the recommended artworks and recommendation reasons. Participants were allowed to pause the experiment and take breaks between sessions as needed. The average duration of the experiment was 90 minutes.

To avoid the effect of the chronological order of the experiment modes on the engagement results, we implemented a counterbalanced design~\cite{pollatsek1995use}. Each participant completed the \textit{Narrative} and \textit{Recommendation sessions} three paintings, with each mode (\textcolor{blue}{\textit{Narrator Mode}}, \textcolor{orange}{\textit{Artist Mode}}, and \textcolor{green}{\textit{In-Situ Mode}}) linking to a different painting in a randomly assigned order. The counterbalanced design ensured that all possible orders were evenly distributed.

\textbf{Onboarding.}
As shown in Figure \ref{Study Process},  in the onboarding session, we first offered a brief introduction about the experiment and the three sessions involved. We then provided participants with training on how to use the prototype. They practiced with an example image and were encouraged to ask questions at any time until they became proficient with the prototype. After that, they completed a pre-survey about demographics and artistic backgrounds.

\textbf{\textit{Narrative Session}.} 
As shown in Figure \ref{Study Process}, participants completed the interactive narrative tasks for four objects under each assigned mode. For each object, participants selected the object, chose the information types, and evaluated the narrative result by think-aloud\cite{van1994think}. Think-aloud is a method where participants verbalize their thoughts and reasoning while performing a task, providing insights into their decision-making process. For the first and third objects, participants could choose four information types (Description, Description + Analysis, Description + Analysis + Interpretation, or Question). For the second and fourth objects, participants chose from five information types, with "Judge" as a mandatory option. This allowed us to simulate scenarios with different information requirements. After completing the interactive narrative task in each assigned mode, they filled out the user engagement scale, followed by the hedonic value scale, epistemic value scale, and perceived character evaluation scale, and then participated in an interview. These measures were used to evaluate their perception of the system and characters, as detailed in the Session \ref{5.4}.

\textbf{\textit{Recommendation Session}.} As shown in Figure \ref{Study Process}, after completing the \textit{Narrative} session in the assigned mode, participants proceeded to the \textit{Recommendation session}, maintaining the same mode used in the \textit{Narrative session}. This session consists of the recommended painting evaluation, recommendation reason evaluation, and character evaluation. First, the character provided two style-oriented recommendations and two object-oriented recommendations and participants rated their satisfaction with each of the four recommended paintings. After clicking each painting, the character provided the reasons for its recommendations. Participants then rated their satisfaction with each recommendation reason and provided a verbal explanation for their ratings by think-aloud. After the satisfaction rating tasks, they filled out the perceived character evaluation scale, specifically for the recommendation task.

\

\subsection{Measures and Data Analysis}\label{5.4}

\subsubsection{\textbf{Narrative Session}}\label{5.4.1}
\

In the \textit{Narrative session}, we used the UES-SF (User Engagement Scale-Short Form)~\cite{o2018practical} to analyze how participants \underline{perceived} the overall engagement level when using different modes of the EyeSee system. This 5-point Likert scale was commonly used to measure the overall perceived engagement across various digital contexts~\cite{bitrian2021enhancing, gabrielli2021engagement, ciotoli2021augmented} and was chosen for its efficiency in reducing survey completion time. This scale was shortened from 31 items to 12 items, comprising Focused Attention (3 items), Perceived Usability (3 items), Aesthetic Appeal (3 items), and Reward Factor (3 items that includes Endurability, Novelty, and Felt Involvement components from the original UES). The complete survey is provided in Appendix A.
We also collected the interaction logs, user-perceived hedonic and epistemic value (5-point Likert scale) \cite{ponsignon2024ability}, and think-aloud data to understand how participants \underline{engaged} with the different characters on the EyeSee system.
First, following previous research~\cite{ben2018investigating}, we collected the task complete time, interaction counts by information types, and satisfaction rating to analyze users' behavioral engagement.  
Second, emotional engagement (hedonic value) was measured using statements like “I had fun with this character” and “This experience was entertaining”. Cognitive engagement (epistemic value) was assessed using statements such as “I learned a lot from this character” and “It was a real learning experience”. The think-aloud data during the recommendation task provided deeper insight into the participants' emotional and cognitive engagement~\cite{ponsignon2024ability}.

After collecting survey and user logs, we performed a one-way ANOVA with a randomized effect to compare the effect of character mode on (1) the overall perceived engagement (2) behavioral engagement based on interaction logs (3) emotional and cognitive engagement based on hedonic and epistemic value survey. A random effect “1/PID” (Participant ID) was included to account for individual differences that could not be explained by the fixed effects in the model~\cite{chang2023citesee}. Post hoc analyses were conducted for pairwise comparisons between the modes. The assumptions for using ANOVA, including normality, homogeneity of variances, and independence of observations, were tested and met. The think-aloud data were analyzed using thematic analysis \cite{braun2006using} to provide insight into emotional and cognitive engagement. The analysis involved several rounds of coding, where three researchers independently reviewed the data and labeled specific segments. After comparison and discussion, they consolidated the different codes into potential overarching themes related to the pros and cons of emotional and cognitive engagement in the three modes. Finally, the researchers independently assigned the final codes to the think-aloud data. Any disagreements during this process were resolved through discussions to ensure consensus. The occurrences of codes were counted and qualitative findings were incorporated in the Section \ref{result1-2-2} and \ref{result1-2-3}.

\subsubsection{\textbf{From the Narrative to Recommendation Session}}
\

To analyze the across-session \underline{evolution} of the character perceptions from the \textit{Narrative} to the \textit{Recommendation session}, we collect the character perceptions based on ~\cite{salminen2024deus}, including (a) consistency, (b) relatability, (c) believability, and (d)stereotypicality.
In the \textit{Recommendation session}, in addition to the character perception scale, we used a 7-point Likert scale to collect satisfaction of recommended paintings and recommendation reasons. 

We performed regression analysis to examine how perceptions in the \textit{Narrative session} influenced the \textit{Recommendation session}. Additionally, we conducted a one-way ANOVA with a randomized effect to analyze the character perception across three modes in two sessions and performed two linear regression analyses to investigate which engagement factors in the \textit{Narrative session} were \underline{associated} with improved satisfaction and perceptions in the \textit{Recommendation session}. 
Specifically, the dependent variables in two sets of regressions were: 1) participants' ratings for art recommendations (i.e., satisfaction ratings of recommended images and recommendation reasons); 2) participants' survey scores on their character perceptions (i.e., consistency, relatability, believability, and stereotypically).
These analyses examined the relationship between the users' behavioral, emotional, and cognitive engagement in the \textit{Narrative session} and their perception of characters and outcomes satisfaction of the recommendations in the \textit{Recommendation session}.

%% file: SECTIONS/8.Appendix8-3.tex

\begin{table*}
\centering
\caption{Participants in the Experience}
\resizebox{0.9\textwidth}{!}{%
\begin{tabular}{@{}ccccccc@{}}
\toprule
\textbf{ID} & \textbf{Gender} & \textbf{Age} & \textbf{Occupation} & \textbf{\begin{tabular}[c]{@{}c@{}}Interested \\ in Art\end{tabular}} & \textbf{\begin{tabular}[c]{@{}c@{}}How often \\ visit Museums\end{tabular}} & \textbf{\begin{tabular}[c]{@{}c@{}}Art as \\ Profession or Hobby\end{tabular}} \\ 
\midrule
P1  & Female & 18  & High school student       & Interested            & Once a month            & Hobby      \\ 
P2  & Female & 18  & High school student       & Interested            & More than once a month  & Hobby      \\ 
P3  & Female & 24  & Undergrad student                     & Very interested       & Several times a year    & Hobby      \\ 
P4  & Female & 23  & Undergrad student                     & Very interested       & More than once a month  & Profession \\ 
P5  & Male   & 32  & AI industry practitioners   & Interested            & Several times a year    & Hobby      \\ 
P6  & Female & 25  & Entrepreneur                & Extremely interested  & More than once a month  & Profession \\ 
P7  & Female & 26  & Art teacher                 & Interested            & Once a month            & Profession \\ 
P8  & Male   & 24  & Undergrad student       & Very interested       & More than once a month  & Hobby      \\ 
P9  & Female & 28  & Master's student                     & Extremely interested  & More than once a month  & Profession \\ 
P10 & Female & 24  & Financial practitioner      & Very interested       & Once a year             & Hobby      \\ 
P11 & Female & 22  & Undergrad student                     & Very interested       & More than once a month  & Profession \\ 
P12 & Male   & 23  & Art archaeology researcher & Very interested       & Several times a year    & Hobby      \\ 
P13 & Female & 24  & Master's student       & Very interested       & More than once a month  & Profession \\ 
P14 & Female & 22  & Freelance artist                  & Extremely interested  & Several times a year    & Hobby      \\ 
P15 & Male   & 20  & Undergrad student                     & Interested            & Several times a year    & Hobby      \\ 
P16 & Female & 25  & Museum staff member    & Interested            & Several times a year    & Profession \\ 
P17 & Female & 31  & Graphic Designer            & Extremely interested  & Several times a year    & Profession \\ 
P18 & Non-binary & 25 & Master's student    & Extremely interested  & Several times a year    & Profession \\ 
P19 & Male   & 24  & Master's student       & Very interested       & Several times a year    & Hobby      \\ 
P20 & Prefer not to say & 23 & Master's student    & Interested            & More than once a month  & Profession \\ 
P21 & Male   & 26  & Master's student                     & Interested            & Several times a year    & Hobby      \\ 
P22 & Male   & 23  & Painter                     & Very interested       & Several times a year    & Hobby      \\ 
P23 & Female & 27  & Master's student                 & Somewhat interested   & Once a year             & Hobby      \\ 
P24 & Male   & 34  & Art teacher                     & Very interested       & Several times a year    & Profession \\ 
\bottomrule
\end{tabular}%
}
\label{tab:participanttable}
\end{table*}

%% file: SECTIONS/6.1.Findings1-1.tex
\section{Results of the \textit{Narrative Session}}

{\textbf{[RQ1]} How do users \underline{perceive} and \underline{engage} with the anthropomorphic characters in three modes-\textbf{\textit{Narrator}}, \textbf{\textit{Artist}}, and \textbf{\textit{In-Situ}}-differently in the \textit{Narrative} \textit{session}}?

Sections \ref{result1-1} and \ref{result1-2} assessed RQ1. Section \ref{result1-1} examined users' overall perceived engagement levels in the three modes of the \textit{EyeSee} system. Section \ref{result1-2} further explained the results of \ref{result1-1} by interpreting the engagement levels through the lenses of behavioral, emotional, and cognitive engagement during user interactions. In these sections, three colors will be used in figures and tables to differentiate the three modes: \textcolor{blue}{\textit{Narrator mode} (blue)}, \textcolor{orange}{\textit{Artist mode} (orange)}, and \textcolor{green}{\textit{In-Situ mode} (green)}.

\subsection{Overall Perceived  Engagement (RQ1a) - Survey 1} \label{result1-1}

\textbf{[RQ1a]} How do users \underline{perceive} the overall engagement level when using three modes in the \textit{Narrative} \textit{session}?

\begin{figure*}
\centering
\includegraphics[width=0.9\textwidth]{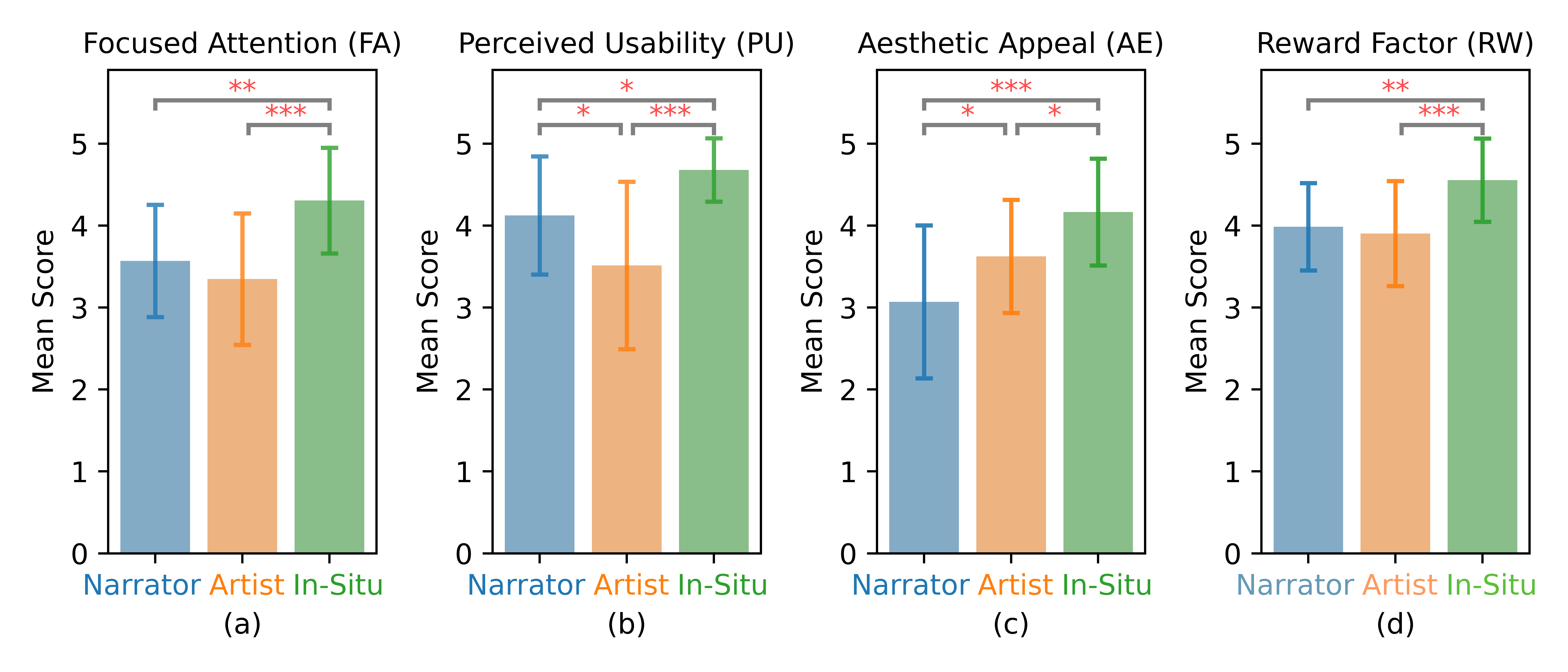}
\caption{\texttt{Survey Results: Overall Perceived Engagement Level}}
\Description{Bar charts showing the mean scores for Focused Attention, Perceived Usability, Aesthetic Appeal, and Reward Factor across three modes (Narrator, Artist, In-Situ), with error bars indicating variability and significant differences marked by asterisks, consistently reveal In-Situ characters as the most engaging across all metrics.}
\label{engagement}
\end{figure*}

In the \textit{Narrative} \textit{session}, participants were asked to rank their engagement level while interacting with three different characters in three modes of the \textit{EyeSee} system. The majority (75\%) selected the \textit{In-Situ mode} as the most engaging, 25\% chose the \textit{Narrator mode} as the most engaging, and 87.5\% considered the \textit{Artist mode} as the least engaging.
As shown in Figure \ref{engagement}, we conducted a one-way ANOVA to assess the overall perceived engagement differences across the three modes, followed by post hoc analyses for pairwise comparisons.
The overall perceived engagement consists of (a) focused attention, (b) perceived usability, (c) aesthetic appeal, and (d) reward factor. Based on the comparison analysis of the survey, significant differences in the user perception of system-related engagement across the three modes were observed.

\textbf{Focused Attention:} There was a significant difference in the users' perception of focused attention when interacting with \textit{EyeSee} across the three modes (\textit{F}(2,51)=11.88, \textit{p}<.001).
As shown in Figure \ref{engagement} (a), participants reported the highest focused attention when interacting with characters in the \textit{In-Situ mode} (M=4.31, SD=0.64); the focused attention in the \textit{Narrator mode} (M=3.57, SD=0.68) was slightly higher than that in the \textit{Artist mode} (M=3.35, SD=0.80). The post hoc analysis revealed that participants exhibited significantly higher focused attention in the \textit{In-Situ mode} than in the \textit{Narrator mode} and \textit{Artist mode}  (\textit{p}<.01 and \textit{p}<.001 respectively), but there was no significant difference in the focused attention level between the \textit{Narrator mode} and \textit{Artist mode}.

\textbf{Perceived Usability:}
There was a significant difference in the users' perception of usability across the three modes (\textit{F}(2,51)=14.32, \textit{p}<.001).
As shown in Figure \ref{engagement} (b), participants reported the highest perceived usability with the \textit{In-Situ mode} (M=4.68, SD=0.39) compared to the \textit{Narrator mode} (M= 4.13, SD=0.72) and the \textit{Artist mode} (M=3.51, SD=1.02). A post hoc analysis showed that participants perceived significantly higher usability in the \textit{In-Situ mode} than the \textit{Narrator mode} and \textit{Artist mode} (\textit{p}<.05 and \textit{p}<.001 respectively). Additionally, participants perceived higher usability in the \textit {Narrator mode} than the \textit{Artist mode} (\textit{p}<.05).

\textbf{Aesthetic Appeal:} There was a significant difference in the users' perception of aesthetic appeal across the three modes (\textit{F}(2,51)=12.24, \textit{p}<.001).
As shown in Figure \ref{engagement} (c), the \textit{In-Situ mode} (M=4.17, SD=0.65) showed the greatest aesthetic appeal, followed by the \textit{Artist mode} (M=3.63, SD=0.69), and the \textit{Narrator mode} (M=3.07, SD=0.93) showed the lowest aesthetic appeal. The post hoc analysis showed that the \textit{In-Situ mode} had significantly greater aesthetic appeal than the \textit{Narrator mode} and \textit{Artist mode} (\textit{p}<.001 and \textit{p}<.05 respectively), and the \textit{Artist mode} had significantly greater aesthetic appeal than the \textit{Narrator mode} (\textit{p}<.05).

\textbf{Reward Factor:}
There was a significant difference in the users' perception of reward factor (\textit{F}(2,51)=9.54, \textit{p}<.001).
As shown in Figure \ref{engagement} (d), the participants reported the highest reward factor when interacting with the characters in the \textit{In-Situ mode} (M=4.56, SD=0.51) on the 5-point Likert scales. The scale included questions assessing whether the experience was "\textit{worthwhile}", "\textit{rewarding}", and "\textit{enjoyable}". The post hoc analysis showed that participants perceived a significantly higher reward factor in the \textit{In-Situ mode} compared to the \textit{Narrator mode} and \textit{Artist mode} (\textit{p}<.01 and \textit{p}<.001 respectively), but there was no significant difference in the reward factor between the \textit{Narrator mode} (M=3.99, SD=0.53) and \textit{Artist mode} (M=3.90, SD=0.64).

\textbf{Summary-RQ1a:}  
According to Figure \ref{engagement}, the \textcolor{green}{\textit{In-Situ mode}} was rated the best user engagement across the four engagement dimensions in terms of focused attention, perceived usability, aesthetic appeal, and reward factor. Between the \textcolor{orange}{\textit{Artist mode}} and \textcolor{blue}{\textit{Narrator mode}}, an interesting contrast was found: the \textcolor{orange}{\textit{Artist mode}} had higher aesthetic appeal but was perceived to have lower usability. 
Next, Section \ref{result1-2} will explain the potential rationale for these results from the behavioral, emotional, and cognitive engagement perspectives.

%% file: SECTIONS/6.1.Findings1-2.tex
\subsection{Users' Behavioral, Emotional and Cognitive Engagement with the Characters (RQ1b) - System Log, Survey 1, Think-Aloud}\label{result1-2}

\textbf{[RQ1b]} How do users  \underline{engage} with different characters in the \textit{Narrative} \textit{session}?

\subsubsection{\textbf{Behavioral Engagement: When Interacting with the \textcolor{green}{In-Situ Mode}, Users' Interaction Time was Longer, and They Asked More Questions Proactively.}}\label{result1-2-1}
\

As shown in Table \ref{perceived engagement}, we investigated the differences in participants' behavioral engagement across the three modes in the \textit{Narrative session}, focusing on interaction time, interaction counts by information type, and satisfaction with the character's responses. First, participants exhibited a significantly longer mean engagement time (29.00 minutes, SD=8.47) when interacting with the \textit{In-Situ mode} compared tothe \textit{Artist mode} (17.96 minutes, SD=7.41) and \textit{Narrator mode} (15.29 minutes, SD=5.15) (\textit{F}(2,51)=24.84, \textit{p}<.001). Furthermore, the \textit{In-Situ mode} evoked the highest number of "Describe" (50), "Analyze" (38), "Interpret" (61), and "Question" (88) behaviors, indicating a more active behavioral engagement. 
Satisfaction ratings showed that the responses of the characters in \textit{In-Situ mode} received the highest approval, with 65\% (SD=0.22) marked as "Like", compared to 59\% (SD=0.27) in the \textit{Narrator mode} and 50\% (SD=0.24) in the \textit{Artist mode}. The percentage of "Neutral" and "Dislike" ratings were also lower in the \textit{In-Situ mode}, further indicating its higher overall satisfaction level. These findings highlighted the effectiveness of \textit{In-Situ mode} in enhancing user behavioral engagement in the \textit{Narrative session}. The results might also explain why the \textit{In-Situ mode} was rated the highest across the four engagement dimensions discussed in section \ref{result1-1}.

\begin{table*}[t!]
\centering
\caption{Behavioral engagement results in the narrative session. Time: Mean (SD). \#: interaction counts by information types.}
\label{tab:RQ1_results}
\begin{tabular}{lccccccccc}
\hline
 & \multicolumn{6}{c}{(i) Art appreciation behaviors based on system log analysis} & \multicolumn{3}{c}{(ii) Satisfaction with output} \\
\cline{2-7} \cline{7-10}
     & \textbf{Time ***} & \textbf{Describe\#} & \textbf{Analyze\#} & \textbf{Interpret\#} & \textbf{Judge\#} & \textbf{Question\#} & \textbf{Like} & \textbf{Neutral} & \textbf{Dislike} \\
\hline
\textbf{Narrator} & {15.29 (5.15)} & 45 & 37 & 59 & 52 & 30 & 59\% & 9\%  & 32\% \\
\textbf{Artist} & 17.96 (7.41) & 44 & 36 & 50 & 57 & 20 & 50\% & 26\%  & 24\% \\
\textbf{In-Situ} & \textbf{29.00 (8.47)} & 50 & 38 & 61 & 50 & \textbf{88} & 65\% & 13\%  & 22\% \\
\hline
\multicolumn{4}{l}{\textit{Note}: * p<0.05, ** p<0.01, *** p<0.001}
\end{tabular}
\label{perceived engagement}
\end{table*}

%% file: SECTIONS/6.1.Findings1-3.tex
\subsubsection{\textbf{Emotional Engagement (EE): \textcolor{green}{In-Situ} Yielded the Highest Hedonic Value, Followed by the \textcolor{orange}{Artist Mode.}}}\label{result1-2-2}
\

\begin{figure}[t!]
\centering
\includegraphics[width=0.45\textwidth]{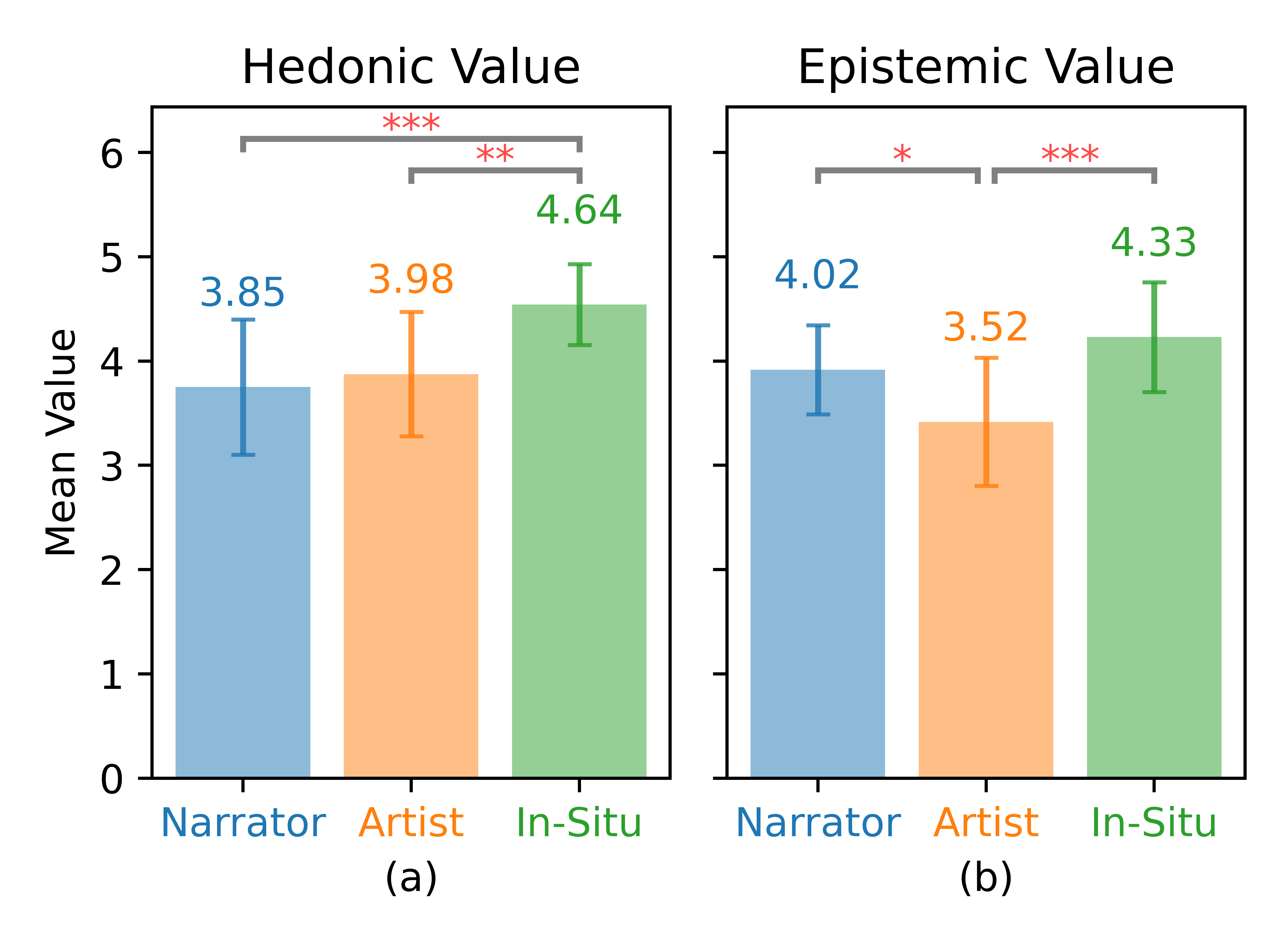}
\caption{\texttt{Hedonic Value and Epistemic Value}}
\Description{Bar charts showing the mean values for Hedonic and Epistemic engagement across three character types (Narrator, Artist, In-Situ), with error bars indicating variability and significant differences marked by asterisks, consistently reveal that In-Situ characters yield the highest values in both engagement types.}
\label{value}
\end{figure}

There was a significant difference in the users' perception of hedonic value across the three modes (\textit{F}(2,51)=8.31, \textit{p}<.001). As shown in Figure \ref{value} (a),
on average, participants rated their emotional engagement as highest in the \textit{In-Situ mode} (M=4.64, SD=0.39) on 5-point Likert scales. A post hoc analysis revealed that emotional engagement was significantly higher in the \textit{In-Situ mode} than the \textit{Artist mode} and \textit{Narrator mode} (\textit{p}<.01 and \textit{p}<.001 respectively). 
While no significant difference was observed between the \textit{Narrator mode} and \textit{Artist mode}, the emotional engagement in the \textit{Artist mode} (M=3.98, SD=0.60) was slightly higher than that in the \textit{Narrator mode} (M=3.85, SD=0.65).
The think-aloud results indicated that the users' emotional engagement in the interactions with the painting object characters (\textit{In-Situ mode}) and \textit{artist characters} (\textit{Artist mode}) was greatly enhanced by the immersive experience involving time travel \textbf{(EE1)}, empathy \textbf{(EE2)}, and anthropomorphism \textbf{(EE3)}. 

\textbf{EE1: \textcolor{orange}{\textit{Artist}} and \textcolor{green}{\textit{In-Situ}}——Time Travel——From Modern to Ancient and From Ancient to Modern.} Time travel in this context refers to the immersive experience where users feel transported between different historical periods during their interactions with the \textit{In-Situ} and \textit{Artist characters}. This experience allows participants to engage with characters from both ancient and modern times, leading a deeper understanding of the cultural and historical context behind the artworks.

\begin{figure*}
    \centering
    \includegraphics[width=0.95\linewidth]{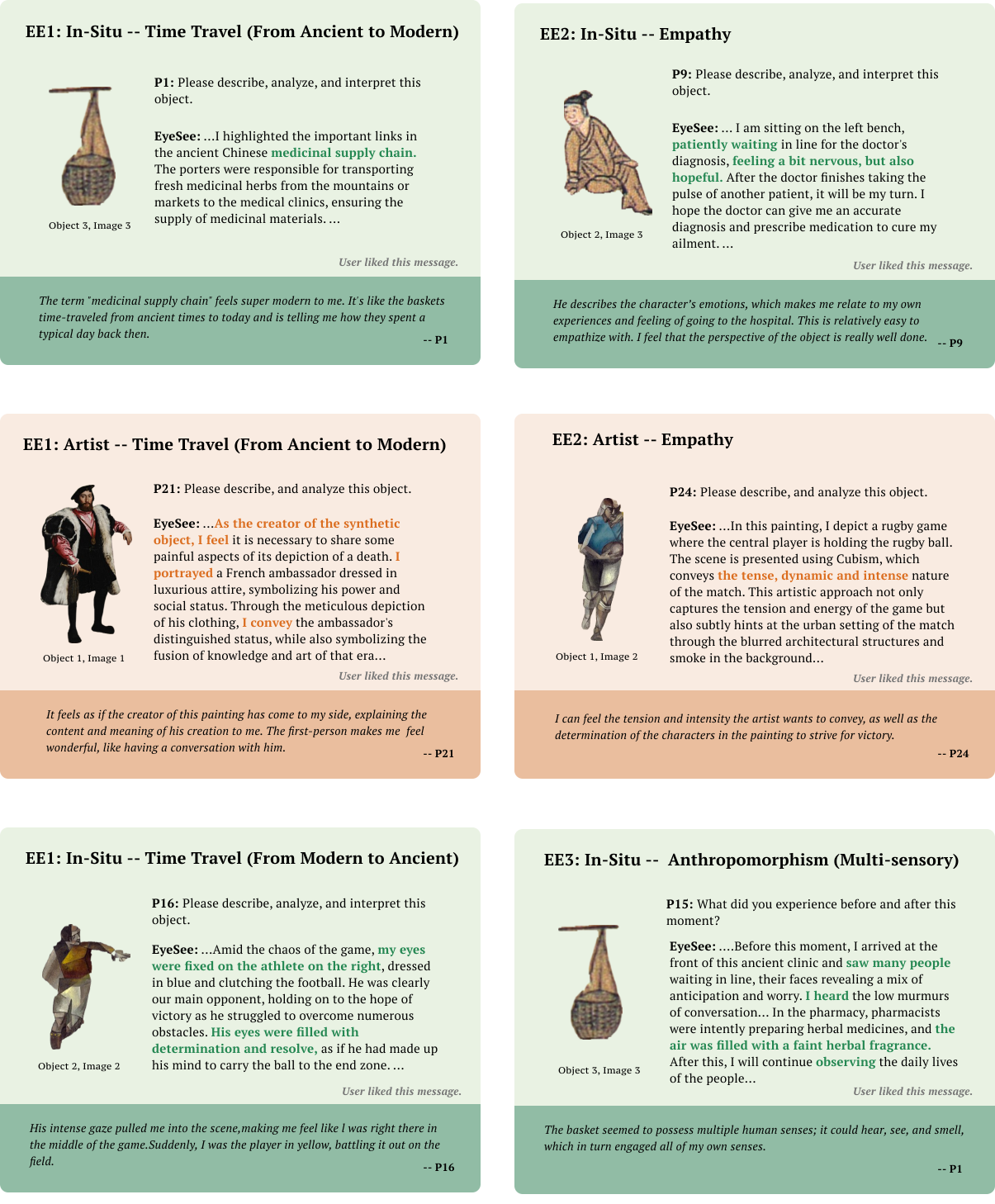}
    \caption{\texttt{Examples of Emotional Engagement}}
    \Description{This image contains six cards, each showcasing user interactions with different objects, including the system’s feedback and the users' comments on that feedback. The cards are categorized by various emotional engagement codes, such as "Time Travel" (from ancient to modern), "Empathy," and "Multi-sensory Anthropomorphism." These cards illustrate how users emotionally engage with objects through the system's interpretation, drawing on personal experiences or reflections. For example, users might feel a connection between ancient objects and modern concepts or feel immersed in the experience through detailed descriptions.}
    \label{emotional}
\end{figure*}

Thirteen participants reported experiencing time travel when interacting with the \textit{In-Situ character}, and four participants reported these experiences with the \textit{Artist character}. Participants expressed that the vivid and detailed narratives created a sense of time travel, transporting them from the modern to the ancient or bringing objects depicted in the painting from the ancient to the modern.
As shown in the examples of EE1 in Figure \ref{emotional}, P1 felt that the term \textit{“medicinal supply chain”} transformed ancient baskets into symbols of historical development, as if they were time travelers narrating the evolution of trade and medicine. P21 highlighted the power of first-person narrative, which made him feel as if he were conversing directly with historical figures or artists. P16 was attracted by the statement \textit{"my eyes were fixed on the athlete on the right"} and felt as though he was an active participant in a modern athletic event. These responses suggested that detailed, first-person storytelling in art appreciation can enhance users' emotional engagement and provide a more profound emotional connection with historical content.

\textbf{EE2:  \textcolor{orange}{\textit{Artist}} and \textcolor{green}{\textit{In-Situ}}——Empathy——Sharing of a Specific Feeling, or Emotion with Character.} 
Empathy, in this context, refers to the emotional connections that participants formed with the characters, where they felt the characters’ emotions, understood their motivations, and expressed care or support during the interaction. The empathy was often triggered by detailed emotional and atmospheric descriptions.

Five participants reported feeling empathy when interacting with the \textit{In-Situ character} and three with the \textit{Artist character}.
As shown in the examples of EE2 in Figure \ref{emotional}, P9 noted that the description of a character’s nervousness and hope while waiting for a doctor’s diagnosis evoked personal memories of hospital visits, leading to a shared emotional experience. Similarly, P24 remarked that the depiction of tension, dynamism, and determination in a cubist-style football game effectively conveyed the artist’s intended emotions, allowing the participant to feel the energy and passion behind the artwork. These empathetic experiences helped deepen participants’ emotional engagement with the characters and the artworks.

\textbf{EE3: \textcolor{green}{\textit{In-Situ}}——Anthropomorphism——Providing a Richer Multi-sensory and Interesting Experience.} 
Anthropomorphism refers to attributing human characteristics to non-human entities. When participants interacted with anthropomorphic \textit{In-Situ characters}, the experience could be multi-sensory. It included hearing, sight, smell, and so on. This multi-sensory experience could enhance the immersion and interestingness in art appreciation. 

Seven participants specifically noted the multi-sensory aspects of their interactions with the \textit{In-Situ character}. Fifteen participants mentioned \textit{"interest"}, \textit{"enjoyment"}, \textit{"fun"}, and \textit{"happiness"} when interacting with the anthropomorphic \textit{In-Situ character}.
As shown in the example of EE3 in Figure \ref{emotional}, P15 appreciated the anthropomorphic multi-sensory narrative that included the \textit{In-Situ character}'s hearing, sight, and smell. Similarly, P16 emphasized the auditory dimension by asking the football, \textit{"In this real environment, according to the laws of physics, what sound can you produce?"}. This illustrated how participants actively engaged with the sensory elements to deepen their connection with the objects. P9 noted the humorous anthropomorphized narrative of the carrying pole, remarking that the description—\textit{"As the carrying pole, I quietly bear the weight of the herbs, contributing my strength to the daily operations of the clinic."}—made the interaction more entertaining. These examples suggested that combining anthropomorphic descriptions with multi-sensory storytelling could significantly enhance emotional engagement, making the experience more immersive and enjoyable.

\textbf{EE4, EE5, EE6: \textit{Three Modes}——Negative Emotion——Dis-appointment, Distrust, and Boredom.} 
While the dialogue with the characters generally aimed to enhance user engagement, some participants experienced negative emotions during their interactions, such as disappointment, distrust, and boredom. 

Disappointment occurred when the characters’ responses did not meet participants’ expectations or failed to provide sufficient information. For example, P7 felt disappointed by the \textit{Narrator character}’s narrative, noting the absence of detailed academic and historical explanations about the evolution of astronomical instruments. Similarly, P18 was disappointed with the \textit{Artist character} for not providing the structural and color analysis of the paintings.

Distrust emerged when participants felt that the characters’ responses were not genuine. For example, P14 expressed skepticism about the \textit{In-Situ character}’s responses, perceiving them as less credible and objective compared to the \textit{Narrator character}. P1 distrusted the \textit{Artist character}, believing that \textit{artists} should avoid subjectively praising their own work, as this limits the viewers’ freedom to interpret and evaluate the art. Three participants also mentioned that the use of bullet points in the \textit{In-Situ mode} made the responses feel less authentic. P19 specifically noted that GPT tends to excel at describing things in bullet points, which can make the \textit{In-Situ characters} less genuine.

Boredom resulted from repetitive information and a lack of novelty in the dialogue. Seven participants reported that repeated content, particularly in the \textit{Artist mode}, made the interactions feel tedious.

%% file: SECTIONS/6.1.Findings1-4.tex
\begin{figure*}
    \centering
    \includegraphics[width=0.95\linewidth]{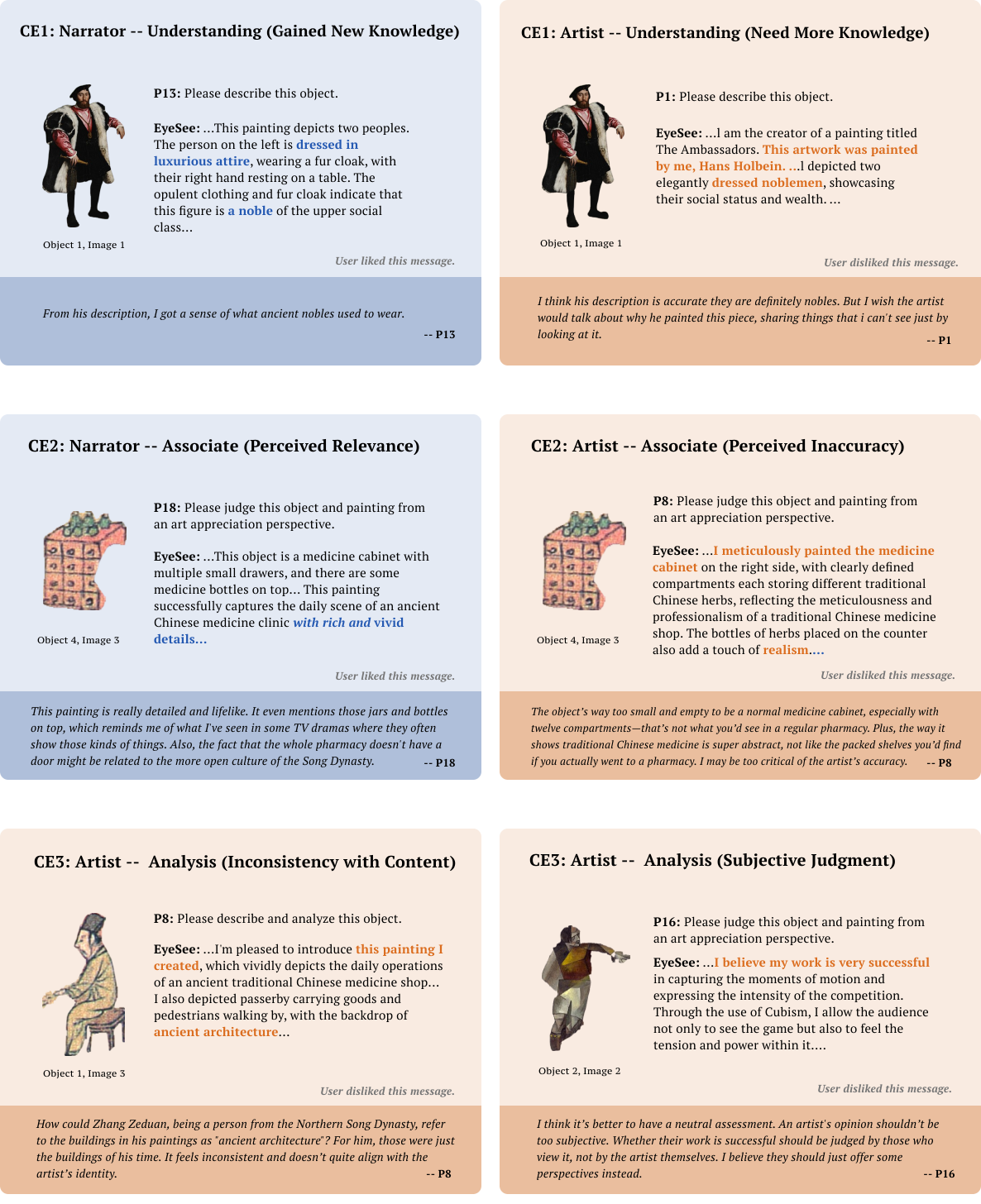}
    \caption{\texttt{Examples of Cognitive Engagement}}
    \Description{This image contains six cards that represent various user interactions with objects, along with system-generated feedback and users' reflections, focusing on different aspects of cognitive engagement. The cards are labeled with cognitive engagement codes, such as "Understanding," "Association," and "Analysis," highlighting how users process and evaluate the information provided. Some cards illustrate users gaining new knowledge about historical objects or understanding the relevance of details, while others show users critically analyzing the system's accuracy or artistic interpretations. Through these interactions, users engage with both the factual and interpretive elements of the objects, sometimes questioning the content, identifying perceived inaccuracies, or making subjective judgments about the artistic choices presented.}
    \label{cognitive}
\end{figure*}

\subsubsection{\textbf{Cognitive Engagement (CE): \textcolor{orange}{\textit{Artist Mode}} Yielded the Lowest Epistemic Value, Compared to the \textcolor{blue}{\textit{Narrator Mode}} and \textcolor{green}{\textit{In-Situ Mode.}}}}\label{result1-2-3}
\

There was a significant difference in the users' perception of epistemic value across the three modes (\textit{F}(2,51)=8.39, \textit{p}<.001). 
As shown in Figure \ref{value} (b), the participants rated their cognitive engagement as lowest in the \textit{Artist mode}. A post hoc analysis revealed that cognitive engagement was significantly lower in the \textit{Artist mode} (M=3.52, SD=0.61) compared to the \textit{Narrator mode} (M=4.02, SD=0.43) and \textit{In-Situ mode} (M=4.33, SD=0.61) (\textit{p}<.05 and \textit{p}<.001 respectively). There was no significant difference in cognitive engagement between the \textit{Narrator mode} and \textit{In-Situ mode}. The think-aloud results provided insight into these findings. First, participants usually had higher expectations for the \textit{Artist characters}, anticipating more understanding of the creative process and background knowledge from the artist's perspective \textbf{(CE1)}. Second, the participants found it unacceptable when the information provided by the \textit{Artist character} was inaccurate \textbf{(CE2)}. Additionally, the participants analyzed the \textit{Artist character} and noted the inconsistencies in content and subjective judgments \textbf{(CE3)}.

\textbf{CE1: \textit{Three Modes}——Understanding——All Provided New Knowledge, but \textcolor{orange}{\textit{Artist}} Needs More.} Participants gained new knowledge and enhanced their understanding of the artwork through interactions with the characters in three modes. When it came to the \textit{Artist character}, participants expected the artist to provide unique insights only the artist knows, such as the creative process or the story behind the work, to deepen their understanding of the painting. 

Twelve participants gained new insights from the \textit{In-Situ character}, and ten from the \textit{Narrator}, while only six did from the \textit{Artist}. Notably, nine participants expressed a desire for the \textit{Artist character} to provide more detailed information. 
As depicted in CE1 in Figure \ref{cognitive}, P13 acquired knowledge about the attire of ancient nobility from the \textit{Narrator character}. Meanwhile, P1 requested more detailed information from the \textit{Artist character} regarding the painting process. Similarly, P9 was interested in learning more about the artist’s creative intentions, and both P12 and P18 found the artist's information to be lacking in specificity. These examples suggested that expanding the knowledge base related to the artist perspective could better align with user expectations.

\textbf{CE2: \textit{Three Modes}——Associate——All Perceived Relevance, but \textcolor{orange}{\textit{Artist}} Noticed Inaccuracy.} 
 Participants associated the information provided by characters with their personal experiences or real-life situations to assess their relevance and accuracy. This association was more prominent in interactions with the \textit{In-Situ} and \textit{Narrator characters}, but less so with the \textit{Artist}, where inaccuracies were more frequently noted.

Nine participants connected information from the \textit{In-Situ character} to their own experiences, while seven did so with the \textit{Narrator} and five with the \textit{Artist}. For example, as illustrated in Figure \ref{cognitive}, P18, using the \textit{Narrator character}, associated the cabinets in TV dramas and agreed with the description \textit{"with rich and vivid details"}. Eight participants noticed inaccuracy while interacting with the \textit{Artist character}. For example, P8 observed that the cabinet in the painting did not match the size of real pharmacy cabinets, which led them to doubt the artist’s claim of realism. This suggested that ensuring accuracy is essential to maintaining trust and engagement with digital characters, particularly in the \textit{Artist mode}.

\textbf{CE3: \textcolor{orange}{\textit{Artist}}——Analysis——Inconsistency with Content and Subjective Judgments.}
Participants identified two key factors that diminished their cognitive engagement with the \textit{Artist character}. First, the information was accurate, but the way it was presented did not align with the artist's historical identity. Second, the artist's evaluations were perceived as subjective.

Six participants noted that the \textit{Artist}’s language and expression did not align with the historical context they expected, while five participants felt that the \textit{Artist}’s judgments were overly subjective. 
As shown in the example of CE3 in Figure \ref{cognitive}, P8 expressed frustration that the \textit{Artist}’s modern language was inconsistent with the tone and style of the period, reducing the authenticity of the interaction. Similarly, P16 argued that the \textit{Artist} should remain neutral and open, avoiding subjectivity. When the \textit{Artist}’s narrative did not match historical expectations or was seen as biased, participants struggled to fully engage with the content, which reduced the epistemic value of the experience.

\textbf{CE4: \textcolor{green}{\textit{In-Situ}}——Curiosity——Active Exploration with \textcolor{green}{\textit{In-Situ}} Characters.}
Participants exhibited strong curiosity when interacting with \textit{In-Situ characters}, actively exploring the content to satisfy their interests.
Eight participants reported this kind of curiosity in the \textit{In-Situ mode}.
For example, P16 asked three consecutive questions about a particular object and expressed a desire to continue asking questions. This active engagement demonstrated that \textit{In-Situ characters} successfully sparked participants' curiosity, promoting a deep understanding of the painting.

\textbf{CE5: \textit{Three Modes}——Reflection——Correcting Characters' Errors.}
Participants recognized and corrected errors in the characters' responses by reflecting and adjusting their interactions. This reflection allowed users to improve the system's understanding of their needs, leading to more accurate information. For instance, they could reselect targets or add contextual information to guide the characters to provide correct interpretation. In one example, when a doctor was mistakenly identified as a patient because the doctor’s stool was not selected, the participant corrected the error by adding the stool to the scene. This action not only resolved the immediate mistake but also demonstrated how users could actively engage in the process. Characters should respond to user interventions and learn from corrections to improve future responses.

\textbf{Summary-RQ1b:} Our findings demonstrated participants' varying levels of engagement with different characters in three modes. First, for behavioral engagement, participants spent significantly longer interaction time and asked more questions when interacting with the characters in the \textcolor{green}{\textit{In-Situ mode}}. Second, regarding emotional engagement, participants experienced the highest hedonic value when interacting with characters in the \textcolor{green}{\textit{In-Situ mode}}, due to immersive experience involving time travel (EE1), anthropomorphism (EE2), and empathy (EE3), followed by the \textcolor{orange}{\textit{Artist mode}}. The emotional engagement explained the aesthetic appeal results in Session \ref{result1-1}. Third, concerning cognitive engagement, participants reported the lowest epistemic value when engaging with the characters in \textcolor{orange}{\textit{Artist mode}} compared the \textcolor{blue}{\textit{Narrator mode}} and \textcolor{green}{\textit{In-Situ mode}}, due to higher knowledge expectations (CE1) and stricter demands for accuracy (CE2) and consistency (CE3), which explained the usability in Session \ref{result1-1}. The Session \ref{result2-1} and \ref{result2-2} explored perception evolution across two sessions.

%% file: SECTIONS/6.2.Findings2-1.tex
\section{Results across the \textit{Narrative} and \textit{Recommendation} \textit{Sessions}}

{\textbf{[RQ2]} How do users perception \underline{evolve} between the \textit{Narrative} and \textit{Recommendation}\textit{ sessions,} and what engagement factors in the \textit{Narrative} \textit{session} are \underline{associated} with the changes?}

Sections \ref{result2-1} and \ref{result2-2} assessed RQ2. Section \ref{result2-1} examined users' perceptions of the characters' responses across two sessions. Section \ref{result2-2} further explained the results of \ref{result2-1} by analyzing the underlying factors in the \textit{Narrative session} that are associated with these perceptions, including behavioral, emotional, and cognitive engagement.

\subsection{The Evolution of Character Response Perception in the Two Sessions (RQ2a)- Survey 1\&2}\label{result2-1}

\

{\textbf{[RQ2a]} How do users' perception of the character's responses \underline{evolve} between the \textit{Narrative} and \textit{Recommendation} \textit{sessions}?}

\begin{figure*}
\centering
\includegraphics[width=0.9\textwidth]{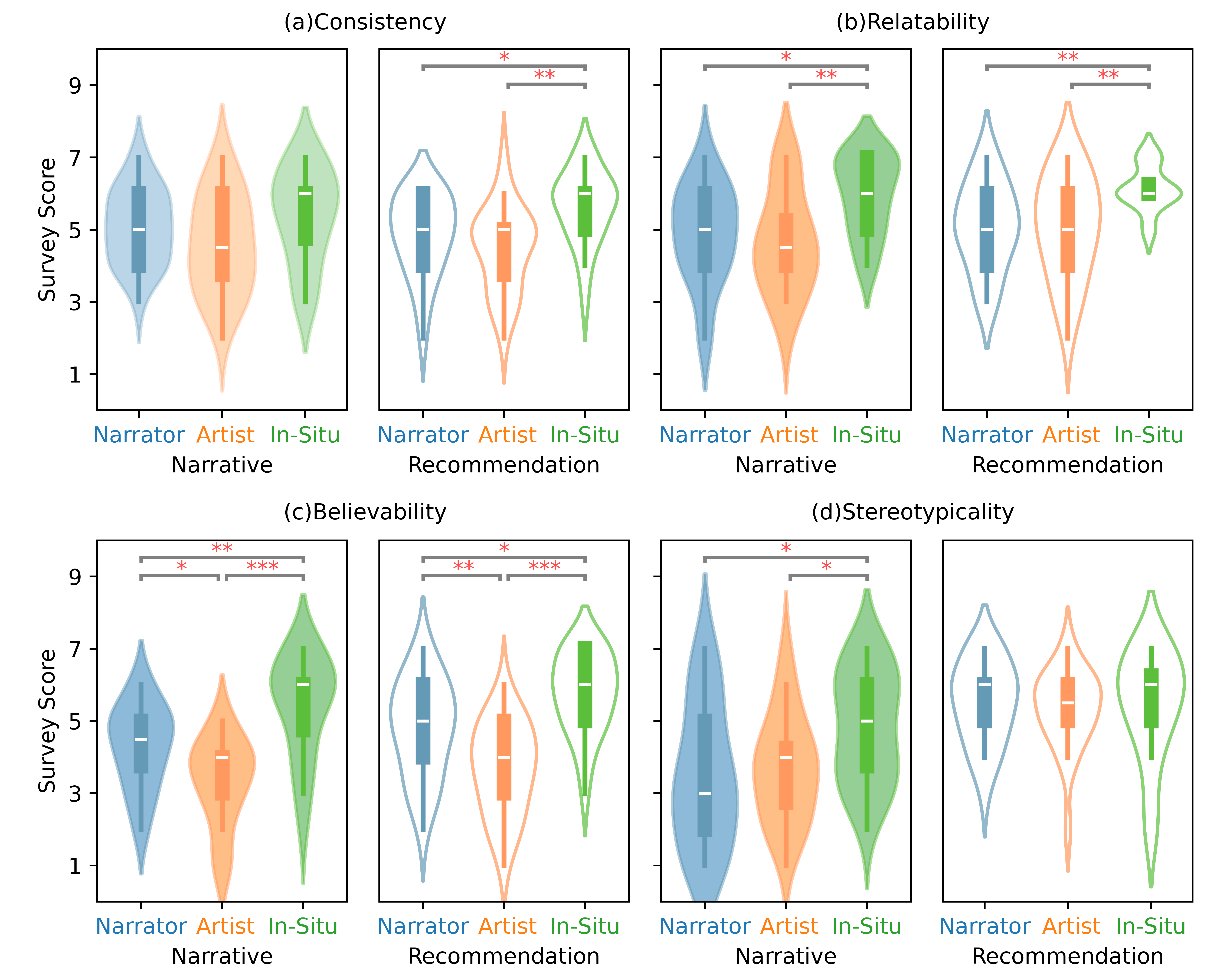}
\caption{\texttt{Survey Results: Users' Perception of the Characters' Responses across Three Modes in Two Sessions.}}
\Description{Violin plots showing survey scores for Consistency, Relatability, Believability, and Stereotypicality across three character types (\textit{Narrator}, \textit{Artist}, \textit{In-Situ}) in both \textit{Narrative} and \textit{Recommendation} contexts, with error bars and significant differences marked by asterisks, indicate that \textit{In-Situ characters} consistently receive the highest scores in all categories except Stereotypicality, where no significant differences were observed.}
\label{recommendation}
\end{figure*}

As shown in Figure \ref{recommendation}, We performed regression analysis to examine how perceptions in the \textit{Narrative session} influenced the \textit{Recommendation session}. Additionally, we conducted a one-way ANOVA with a randomized effect to analyze the character perception across three modes in two sessions, followed by post hoc analyses for pairwise comparisons. The perception of characters consists of (a) consistency, (b) relatability, (c) believability, and (d) stereotypicality. Based on comparison results from the survey, significant differences in perceptions of character were observed when interacting with the three types of characters across the \textit{Narrative} and \textit{Recommendation} \textit{sessions}.

\textbf{Consistency:}
The regression analysis showed no correlation between the perceived response consistency in the \textit{Narrative} \textit{session} and the \textit{Recommendation} \textit{session} ($\beta$=0.17, S.E.=0.11), suggesting that users' perceptions of information consistency may vary independently across the two sessions. There was no significant difference in perceived character response consistency across the three modes in the \textit{Narrative} \textit{session}. In contrast, in the \textit{Recommendation} \textit{session}, there was a significant difference in the perceived character response consistency (\textit{F}(2,51)=6.92, \textit{p}<.01). The post hoc analysis found that the perceived response consistency of the \textit{In-Situ characters} (M=5.63, SD=1.01) was significantly higher than that of the \textit{Narrator characters} (M=4.83, SD=1.13) and \textit{Artist characters} (M=4.46, SD=1.18) (\textit{p}<.05 and \textit{p}<.01 respectively) in the \textit{Recommendation} \textit{session}. The median perceived response consistency in the \textit{Recommendation} \textit{session} (median=5.0) was higher than that in the \textit{Narrative} \textit{session} (median=4.5).

\textbf{Relatability:}
The regression analysis showed a significant correlation between the perceived response relatability in the \textit{Narrative session} and the \textit{Recommendation session} ($\beta$=0.30, S.E.=0.10, \textit{p}<0.01), indicating that perceived relatability in \textit{Narrative session} had a positive effect on that in the \textit{Recommendation session}.
In the \textit{Narrative session}, there was a significant difference in perceived characters' responses relatability across the three modes (\textit{F}(2,51)=6.81, \textit{p}<.01). Similarly, the \textit{Recommendation session} showed a significant difference in perceived characters' responses relatability (\textit{F}(2,51)=6.98, \textit{p}<.01).
The post hoc analysis found that \textit{In-Situ characters'} (M=6.13, SD=0.61) responses had significantly higher perceived reliability than \textit{Narrator characters'} (M=5.08, SD=1.21) and \textit{Artist characters'} (M=5.04, SD=1.43) (both \textit{p}<.05). The median perceived characters' responses relatability in the \textit{Recommendation session} (median=5.0) was higher than that in the \textit{Narrative session} (median=4.5).

\textbf{Believability:}
The regression analysis showed a significant correlation between the perceived response believability in the \textit{Narrative session} and the \textit{Recommendation session} ($\beta$=0.45, S.E.=0.11, \textit{p}<0.001), indicating that perceived believability in \textit{Narrative session} had a positive effect that in the \textit{Recommendation session}.
There was a significant difference in perceived characters' responses believability across the three modes in the \textit{Narrative session} (\textit{F}(2,51)=16.35, \textit{p}<.001). 
In the \textit{Recommendation session}, there was also a significant difference in perceived characters' responses believability, with similar trends observed across all modes (\textit{F}(2,51)=16.71, \textit{p}<.001).
The post hoc analysis showed that \textit{In-Situ characters'} (M=5.75, SD=1.11) responses had significantly greater believability than \textit{Narrator characters'} (M=4.79, SD=1.35) and \textit{Artist characters'} (M=3.67, SD=1.27) (\textit{p}<.05 and \textit{p}<.001 respectively). Additionally, \textit {Narrator characters'} responses were perceived as significantly more believable than \textit{Artist characters'} (\textit{p}<.01). 

\textbf{Stereotypicality:} 
The regression analysis showed no correlation between the perceived response stereotypicality in the \textit{Narrative session} and the \textit{Recommendation session} ($\beta$=-0.06, S.E.=0.08).
In the \textit{Narrative session}, there was a significant difference in perceived characters' responses stereotypicality across the three modes (F(2,51)=4.67, \textit{p}<.05). However,
In the \textit{Recommendation session}, there was no significant difference in perceived characters' responses stereotypicality.  
The lack of significant differences in the \textit{Recommendation session} suggested that participants might have focused more on the practical content of the characters’ recommendations, rather than their stereotypical traits. In contrast, the open-ended nature of the Narrative task may have led participants to perceive some characters as conforming to stereotypes, as their responses were less structured.

\textbf{Summary-RQ2a:} 
We compared participants' perceptions of the characters' responses across three modes in the two sessions. First, the \textcolor{green}{\textit{In-Situ characters}} consistently received the highest scores in consistency, relatability, and believability in both interaction sessions (i.e., \textit{Narrative} and \textit{Recommendation}). Second, we observed the gap in the perceived consistency between the \textcolor{orange}{\textit{Artist character}} and the \textcolor{blue}{\textit{Narrator character}} narrowed in the \textit{Recommendation session} and the perceived stereotypicality significantly changed across the two sessions. Next, Session \ref{result2-2} explained the potential rationale for these results from interactive engagement perspectives.

%% file: SECTIONS/6.2.Findings2-2.tex
\subsection{Across-session Association of Engagement Factors in the \textit{Narrative Session} (RQ2b) - System Log, Survey 1\&2, Think-Aloud} \label{result2-2}

\textbf{[RQ2b]} What engagement factors in the \textit{Narrative Session} are \underline{associated} with the user perception in the \textit{Recommendation session}?

To address this research question, we performed two linear regression analyses to explore how engagement factors in the \textit{Narrative Session} influenced participants' character perception and outcomes satisfaction in the \textit{Recommendation session}.
The two sets of regressions aim to yield insights about users’ perceptions from two perspectives: 1) the first set of regressions used participants’ character response perception survey scores (i.e., 
consistency, relatability, believability, and stereotypicality) as the dependent variables; 2) the second set of regressions adopted participants’ satisfaction ratings for recommendation images and reasons as dependent variables respectively. Based on our previous findings, we used factors that could represent users’ interactive engagement as predictors: behavioral engagement (measured by interaction time), emotional engagement (measured by hedonic value), and cognitive engagement (measured by epistemic value). As discussed in Sections \ref{result1-2-1}, \ref{result1-2-2}, and \ref{result1-2-3}, these factors differ significantly across different characters.

\begin{table*}[t!]
\centering
\caption{Regression results with users’ engagement data in the narrative session as predictors and perceived character survey ratings in the recommendation session as dependent variables. Each column in the table represents one regression performed with the corresponding rating item as the dependent variable.}
\begin{tabular}{p{3cm}p{2.0cm}p{2.0cm}p{2.0cm}p{2.0cm}}
\hline
& \multicolumn{4}{c}{\textbf{Dependent Variables — Experience (Survey Scores)}} \\
\cline{2-5}
\textbf{Predictors} 
& \multicolumn{1}{c}{\textbf{Consistency}} 
& \multicolumn{1}{c}{\textbf{Relatability}}
& \multicolumn{1}{c}{\textbf{Believability}}
& \multicolumn{1}{c}{\textbf{Stereotypicality}} \\
& \multicolumn{1}{c}{\textbf{$\beta$(S.E.)}}
& \multicolumn{1}{c}{\textbf{$\beta$(S.E.)}}
& \multicolumn{1}{c}{\textbf{$\beta$(S.E.)}}
& \multicolumn{1}{c}{\textbf{$\beta$(S.E.)}}\\
\hline

Behavioral 
Engagement & \centering{0.01 (0.01)} & \centering{0.03 (0.02)} & \centering{0.01 (0.01)} &  \multicolumn{1}{c}{0.04 (0.02)*} \\
Emotional 
Engagement & \centering{0.32 (0.18)} & \centering{0.14 (0.19)} & \centering{0.74 (0.14)***}  & \multicolumn{1}{c}{-0.13 (0.20)} \\
Cognitive 
Engagement & \centering{0.52 (0.18)**} & \centering{0.48 (0.19)*} & \centering{0.53 (0.14)***}  & \multicolumn{1}{c}{0.01 (0.21)} \\

\hline

\multicolumn{4}{l}{\textit{Note}: * \textit{p}<0.05, ** \textit{p}<0.01, *** \textit{p}<0.001}
\end{tabular}
\label{regression2}
\end{table*}

\subsubsection{\textbf{Perceived consistency in the recommendation session was associated with cognitive engagement in the narrative session and stereotypicality was associated with behavioral engagement.}} Table \ref{regression2} showed the regression results with survey scores of characters' responses perception in the \textit{Recommendation session} as dependent variables and interactive engagement factors in the \textit{Narrative session} as predictors.

For consistency and relatability, we found a significant positive association between participants' cognitive engagement in the \textit{Narrative session} and their perceived consistency($\beta$=0.52, S.E.=0.18, \textit{p}<0.01) and relatability ($\beta$=0.48, S.E.=0.19, \textit{p}<0.05) of the character's responses in the \textit{Recommendation session}. This suggested that when participants experience a higher level of cognitive engagement in the \textit{Narrative session}, they tend to perceive the character responses as more consistent and relatable during the \textit{Recommendation session}. Participants’ reflections during the think-aloud process supported these findings. For example, one comment from P20 was \textit{"Because I felt his answers were good in the previous chat stage, I think his recommendations are consistent with his identity."}. 

For believability, we found a significant positive association between participants' emotional engagement ($\beta$=0.74, S.E.=0.14, \textit{p}<0.001) and cognitive engagement ($\beta$=0.53, S.E.=0.14, \textit{p}<0.001) in the \textit{Narrative session} and their perceived believability for the character's responses in the \textit{Recommendation session}. This indicated that both emotional and cognitive engagement in the \textit{Narrative session} played critical roles in shaping participants' trust in the character's recommendations.  Participants also reflected on this during their think-aloud process. For example, P19 illustrated this sentiment, stating \textit{"Generally, when I use these characters, if they give me incorrect analysis, I will tolerate it and use them a second time. But if they are wrong again, I will never use them again and won't trust their recommendations anymore."}. No statistically significant association was found between behavioral engagement ($\beta$=0.01, S.E.=0.01) and believability.

Third, stereotypicality was found to be associated with interaction time ($\beta$=0.04, S.E.=0.02, \textit{p}<0.05), indicating that the brevity of interactions in the \textit{Recommendation session} may not have provided sufficient time for users to perceive differences in stereotypicality. Participant reflections support this finding. For example, P24 noted, \textit{“I didn’t really notice much difference in how stereotypical the characters were during the short recommendation task, but when I spent more time with them in the \textit{Narrative} task, those differences became more obvious.”}.

\begin{table}[t!]
\centering
\caption{Regression Results with outcomes satisfaction ratings as dependent variables and users’ engagement data as predictors.}
\begin{tabular}{p{3cm}p{2cm}p{2cm}}
\hline
& \multicolumn{2}{c}{\textbf{Dependent Variables — Outcomes}} \\
& \multicolumn{2}{c}{\textbf{(Recommendation Ratings)}}\\
\cline{2-3}
\textbf{Predictors}
& \multicolumn{1}{c}{\textbf{Image}}
& \multicolumn{1}{c}{\textbf{Reasons}} \\
& \multicolumn{1}{c}{\textbf{Satisfaction}}
& \multicolumn{1}{c}{\textbf{Satisfaction}} \\
& \multicolumn{1}{c}{\textbf{$\beta$(S.E.)}}
& \multicolumn{1}{c}{\textbf{$\beta$(S.E.)}}\\
\hline
Behavioral Engagement & \centering{0.03 (0.01)***} & \multicolumn{1}{c}{0.03 (0.02)} \\
Emotional Engagement & \centering{0.39 (0.10)***} & \multicolumn{1}{c}{-0.03 (0.22)} \\
Cognitive Engagement & \centering{0.53 (0.10)***} & \multicolumn{1}{c}{0.36 (0.23)} \\
\hline
\multicolumn{3}{l}{\textit{Note}: * \textit{p}<0.05, ** \textit{p}<0.01, *** \textit{p}<0.001}
\end{tabular}
\label{regression1}
\end{table}

\subsubsection{\textbf{Satisfactions with generated recommended images were associated with engagement factors in the \textit{Narrative session}.}} As shown in Table \ref{regression1}, a positive association was identified between recommended image satisfaction ratings in the \textit{Recommandation session} and behavioral engagement in the \textit{Narrative session} ($\beta$=0.03, S.E.=0.01, \textit{p}<0.001). Additionally, emotional engagement ($\beta$=0.39, S.E.=0.10, \textit{p}<0.001) and cognitive engagement ($\beta$=0.53, S.E.=0.10, \textit{p}<0.001) in the \textit{Narrative session} were also significantly associated with satisfaction ratings for the recommendation images. These findings suggested that participants who were more engaged during the \textit{Narrative session} perceived the outcome of the \textit{Recommendation session} as higher in quality.

In contrast, satisfaction with recommendation reasons showed weaker associations with engagement factors, and all were non-significant. This suggested that participants' engagement in the \textit{Narrative session} did not influence how satisfied they were with the explanations given for the recommendations.

Participants provided additional insight into these findings during their think-aloud process. Some participants, such as P3, P6, P16, and P19, expressed a preference for straightforward, factual explanations—whether object-based, stylistic, or author-centric—indicating that they did not require detailed justifications. This feedback suggested that for some users, a deep familiarity with art might reduced the need for recommendation reasons. However, other participants noted that the explanations were often too superficial, citing descriptions like "the two paintings share the same style" as overly generic and uninformative. These superficial explanations could potentially erode trust when factual inaccuracies were present.

Participants also suggested that the effectiveness of the recommendations could be improved by aligning the types of recommendations with specific characters. For example, the \textit{Artist character} could present recommendations based on the same artist, while Narrator could deliver composition-based recommendations. \textit{In-Situ character} could introduce Object-specific recommendations. This feedback highlighted the importance of character differentiation in enhancing the relevance and appeal of the recommendations.

\textbf{Summary-RQ2b:} Our findings demonstrated participants' engagement in the \textit{Narrative session} was associated with their perceptions of character responses in the \textit{Recommendation session}. Specifically, we identified three key associations between engagement factors and character response perception. First, cognitive engagement in the \textit{Narrative session} was positively associated with perceived consistency in the \textit{Recommendation session}. This helped explain the improvement in the perceived response consistency of \textcolor{orange}{\textit{Artist character}} in Session \ref{result2-1}, as users who were more cognitively engaged during the \textit{Narrative session} likely viewed the character’s recommendations as more coherent and aligned. Second, both emotional and cognitive engagement in the \textit{Narrative session} was related to believability in the \textit{Recommendation session}. Third, interaction time was associated with stereotypicality, which suggested the brief recommendation interactions were not sufficient for users to perceive differences in stereotypicality of \ref{result2-1} between characters.

%% file: SECTIONS/7.Discussion.tex
\section{DISCUSSION}

A major contribution of our work is to demonstrate that multi-character interaction--\textcolor{blue}{\textit{Narrator}}, \textcolor{orange}{\textit{Artist}}, and \textcolor{green}{\textit{In-Situ}}--in art appreciation systems can enhance user engagement in the \textit{Narrative session}. These engagement factors in the \textit{Narrative session} were associated with participants' perceptions of the character responses in the \textit{Recommendation session}. By fostering meaningful interactions using first-person anthropomorphic narratives, the characters shaped users' perceptions (i.e. satisfaction, reliability, and trust) of their recommendations in the \textit{Recommendation session}. In turn, users' expectations of characters' ability to complete tasks and the fitness between characters and tasks will also shape users' perception of characters.

\subsection{First-Person Perspective: A Double-Edged Sword for User Engagement and Critical Response}

Regarding RQ 1, which investigates how users \underline{perceive} and \underline{engage} with anthropomorphic characters in three modes differently in the \textit{Narrative session}, our findings suggest that the overall perceived engagement in terms of aesthetic appeal and perceived usability varied across the three modes. This variation is further interpreted through the lenses of behavioral, emotional, and cognitive engagement during user interactions.

For aesthetic appeal, both the \textit{In-Situ} and \textit{Artist modes} were rated higher than the \textit{Narrator mode}. This aligns with previous research~\cite{busselle2009measuring, salem2017does, samur2021getting, brennan2024versus}
, which suggested that first-person narratives in interactive systems can effectively enhance the aesthetic appeal of the systems. Our study contributes to the existing literature by demonstrating that users' emotional engagement play a key role in shaping their perception of system aesthetic appeal. Participants in the first-person \textit{In-Situ mode} displayed the highest level of interaction time, posed the most questions, and experienced the greatest hedonic value. This immersive mode innovated the perspective and identity through which information was presented, allowing participants to experience elements like time travel (EE1), empathy (EE2), and anthropomorphism (EE3) without a need for an expanded knowledge base. These features contributed to a sense of immersion, suggesting that future \textit{In-Situ characters} equipped with a more robust knowledge base might yield even greater engagement.

Interestingly, \textit{Artist mode} was scored lower on perceived usability compared to \textit{Narrator} and \textit{In-Situ mode}. The findings suggest that the character's identity in this mode plays a crucial role in shaping the perceived usability. Two key cognitive factors may explain this. First, users often expect the artist to provide unique insights only the artist knows, such as the details about the artwork's creative process. According to the Expectation-Confirmation Model (ECM)\cite{bhattacherjee2001understanding}, when users’ high expectations are not fully met, the high expectations might lead to lower perceived usefulness and usability, especially through first-person narratives. Second, users tend to view the artist as a professional and authority, meaning their tolerance for mistakes or inconsistencies of such character is lower. This could be explained using Authority Theory~\cite{pace2006classroom, komorowska2021first}, which suggested that when a person makes statements about their intentions or beliefs, others might be more inclined to question or be less certain about the authority of those statements compared to statements about inner phenomenal states (such as sensations and emotions). Therefore, we suggest future research to deploy different types of experts in the diverse education settings, to explore the impact of potential expert character information on users' use of the system.

\subsection{Influence of Task Type on Role Perception: Unpacking Consistency and Stereotypes}

Regarding RQ2, which explores how users \underline{perception} evolve between \textit{Narrative} and \textit{Recommendation session}s and the across-session \underline{association} of the engagement factors in the \textit{Narrative session}, our results suggest that tasks types had a significant influence on users' character perception (i.e., consistency and stereotypicality) across art \textit{Narrative session} and art \textit{Recommendation session}.

For consistency, the \textit{In-Situ characters} consistently achieved the highest scores across both sessions. In contrast, the response consistency score of the \textit{Artist character} improved from the \textit{Narrator session} to \textit{Recommendation session}, which might also be explained by the Expectation-Confirmation Model (ECM)~\cite{bhattacherjee2001understanding}. First, because the \textit{Artist character} provided new epistemic values aligned with users' expectations during the \textit{Recommendation session} and there was a correlation between users' perception of the characters' response consistency and their cognitive engagement during interactions. Secondly, users expressed that it was natural and appropriate for the \textit{Artist character} to recommend their works, related artworks, or those from the same period. They found it seems appropriate for \textit{In-Situ characters} to provide object-based recommendations. This expands the concept of character consistency error described by Welleck et al.~\cite{welleck2018dialogue}, which defines a consistency error as an utterance unlikely to be made by a character defined by a specific set of traits. Here, users' perceptions of character consistency encompass not just the relationship between the character and the information, but also the consistency between the character's role and the task at hand. Third, these shifts suggest that as users moved from a somehow subjective task to an objective task, their expectations of character responses shifted, particularly regarding how consistent the character responses were perceived in their recommendations.

For stereotypicality, we found that the \textit{In-Situ characters} showed lower character stereotype in the \textit{Narrative session}, but there was no statistical difference between the three character stereotype perceptions in the \textit{Recommendation session}. According to the regression analysis results, users' perception of character stereotypes was associated with their behavioral engagement (interaction time) during interactions. This could be due to the fact that \textit{Narrative session} involved a prolonged, multi-interaction process, whereas the \textit{Recommendation} involved a single interaction. This finding aligns with previous research findings~\cite{lee2020hear}, that interaction time significantly affects the relationship building between humans and chatbots. Furthermore, according to \ref{result1-2-1}, users interacting with \textit{In-Situ characters} were more willing to ask questions because they felt surprised as the \textit{In-Situ character} provided unexpected information during multiple narrative interactions, that broke the stereotype. This is consistent with observations of Ha et al.~\cite{ha2024clochat}, who noted that conversations tended to be longer when users found them more engaging. Thus, these results suggest that stereotypes can be mitigated by enhancing engagement and prolonging interaction time.

\subsection{Design Implications}

Based on our discussions, we propose the following design implications for art appreciation and diverse education settings.

\subsubsection{\textbf{Art Appreciation}}
\

\textbf{Leverage \textit{In-Situ Character} Interactions to Enhance Art Appreciation}
While previous research has proposed various mechanisms behind the benefits of art appreciation~\cite{fancourt2019evidence, mastandrea2019art}, experimental evidence remains limited~\cite{trupp2023benefits}. Our findings demonstrate the unique potential of \textit{In-Situ character} interactions in enhancing user engagement, extending beyond traditional digital technologies. While short art appreciation sessions have been proved to offer mental and physical benefits for users~\cite{clow2006normalisation, mastandrea2019visits, ho2015art}, \textit{EyeSee} allows for more sustained aesthetic experiences, with characters serving as novel mediators of artistic communication~\cite{dervin2023critical, silva2006distinction}.

\textbf{Expand Virtual Art Spaces with Immersive \textit{In-Situ Character} Design.}
Incorporating anthropomorphic characters in digital art systems not only enhances user engagement but also introduces a new interaction medium for the artistic community. Wang et al.~\cite{wang2024virtuwander} introduced VirtuWander, a voice-controlled virtual museum prototype that highlights five interaction designs (Voice, Avatar, Text Window, Highlight, Virtual Screen), demonstrating the value of single-character systems. Adding \textit{In-Situ characters} in virtual spaces can further enrich the design space and provide a deeper and more immersive art appreciation experience.

\subsubsection{\textbf{Diverse Educational Settings}}

\

\textbf{Prioritize \textit{In-Situ Mode} to Enhance User Engagement and Trust.}
Meta-analyses show agents have a small, positive effect on learning~\cite{schroeder2013effective}, but a systematic review found no significant differences~\cite{heidig2011pedagogical}. Our results indicate that emotional engagement in the \textit{In-Situ mode} was significantly higher than the other two modes in art appreciation.  suggesting future systems could prioritize this mode to analyze the impact of \textit{In-Situ mode} on learning outcomes. Additionally, we found this emotional engagement correlated strongly with perceptions of believability in the \textit{Recommendation session}. This inspires researchers to focus on relationship building through interactive recommendation in learning environments where virtual teachers are required to provide advice, to improve users' satisfaction and trust in information.

\textbf{Use the First-person Sparingly in Authority Roles.}
First-person characters, while engaging, may be perceived as overly authoritative, negatively impacting system usability. Research shows a tension between authoritative and dialogic approaches, as the introduction of new ideas is authoritative to support learning~\cite{scott2006tension}. For authority figures, using first-person sparingly could prevent this tension~\cite{marsh2014improving}. 
This has inspired the use of character perspective in some online learning scenarios that may lead users to develop a sense of authority.

\textbf{Make the Character Adaptive (Objective and Subjective).}
Our findings show that characters are perceived differently across tasks, suggesting that adaptive algorithms should adjust character behavior based on both the learning goals and the context in which learning occurs. Subjective and objective education emphasize different values~\cite{sanasintani2024philosophical}. Therefore, combining our conclusions about hedonic and epistemic values can inspire research on using characters in two relatively distinct disciplines, such as STEM and history.

\subsection{Limitations and Future Research}

Our study, while shedding light on the diverse user art appreciation experiences with diverse characters in LLMs, has several limitations that must be acknowledged. Firstly, our participant pool consisted exclusively of art enthusiasts, which may influence the generalizability of our findings to broader populations. Future studies should include participants with varying levels of art interest to broaden the applicability of the results. Secondly, the characters creation on the \textit{EyeSee} system relied on prompt injection, which may lack depth in specialized domains \cite{ankush2023kitlm}. Future research could explore fine-tuning techniques \cite{nguyen2024information, shrestha2024post} or external memory integration \cite{salminen2018persona, ha2024clochat} to improve character customization and enhance personalized art experience. Thirdly, personalization of the art experience could be enhanced by allowing users to select artwork they are interested in, offering greater flexibility and personalization. Future work should also explore alternative interaction methods, such as mobile apps or VR, to improve engagement. Lastly, we did not examine long-term user experiences with \textit{EyeSee}. Future research should conduct longitudinal studies to explore how user interactions evolve over time, providing deeper insights into sustained engagement with conversational agents like \textit{EyeSee}.